\documentclass[aps,prd,onecolumn,10pt,amsfonts,amssymb,amsmath,nofootinbib,showpacs,notitlepage]{revtex4}

\usepackage{graphicx}
\usepackage{bm}

\def\be{\begin{equation}}
\def\ee{\end{equation}}
\def\ba{\begin{eqnarray}}
\def\ea{\end{eqnarray}}
\def\bs{\begin{subequations}}
\def\es{\end{subequations}}
\def\C{{\cal C}}

\newcommand{\Mpl}{M_{\rm pl}}

\begin{document}

\title{Primordial non-Gaussianities in general modified gravitational models
of inflation}

\author{Antonio De Felice}
\affiliation{Department of Physics, Faculty of Science, Tokyo University of Science,
1-3, Kagurazaka, Shinjuku-ku, Tokyo 162-8601, Japan}

\author{Shinji Tsujikawa}
\affiliation{Department of Physics, Faculty of Science, Tokyo University of Science,
1-3, Kagurazaka, Shinjuku-ku, Tokyo 162-8601, Japan}

\begin{abstract}
We compute the three-point correlation function of primordial scalar
density perturbations in a general single-field inflationary scenario,
where a scalar field $\phi$ has a direct coupling with the Ricci scalar
$R$ and the Gauss-Bonnet term ${\cal G}$.
Our analysis also covers the models in which the Lagrangian includes a
function non-linear in the field kinetic energy $X=-(\partial \phi)^2/2$,
and a Galileon-type field self-interaction $G(\phi, X)\square \phi$, 
where $G$ is a function of $\phi$ and $X$. 
We provide a general analytic formula for the equilateral 
non-Gaussianity parameter $f_{\rm NL}^{\rm equil}$ associated with 
the bispectrum of curvature perturbations.
A quasi de Sitter approximation in terms of slow-variation
parameters allows us to derive a simplified form of 
$f_{\rm NL}^{\rm equil}$ convenient to constrain 
various inflation models observationally.
If the propagation speed of the scalar perturbations is much smaller 
than the speed of light, the Gauss-Bonnet term as well as 
the Galileon-type field self-interaction
can give rise to large non-Gaussianities testable in future observations.
We also show that, in Brans-Dicke theory with a field potential 
(including $f(R)$ gravity), $f_{\rm NL}^{\rm equil}$ is 
of the order of slow-roll parameters as in standard inflation 
driven by a minimally coupled scalar field.
\end{abstract}

\date{\today}


\maketitle

\section{Introduction}

The idea of cosmic acceleration in the early 
Universe \cite{Sta80,oldinf} was originally 
introduced as a way of addressing the
flatness and horizon problems plagued in the big bang 
cosmology. The inflationary paradigm can provide a causal 
mechanism for generating density perturbations responsible 
for the Cosmic Microwave Background (CMB) temperature
anisotropies \cite{oldper}. The standard, single-field slow-roll models 
of inflation predict adiabatic and Gaussian primordial 
perturbations with a nearly scale-invariant spectrum.
This property is consistent with the observed CMB 
anisotropies \cite{COBE,WMAP1and5,WMAP7} 
as well as the large-scale structure data \cite{LSS}. 

Over the past 30 years, many models of inflation have been 
proposed in the framework of particle physics or 
extended theories of gravity (see e.g., \cite{review} for reviews).
{}From the information of the spectral index $n_{\cal R}$
of the scalar perturbations as well as the tensor-to-scalar 
ratio $r$, the CMB observations by the WMAP satellite
have been able to rule out some of those 
models \cite{WMAP1and5,WMAP7,alawmap}. 
While the observables $n_{\cal R}$ and $r$ are derived in linear cosmological 
perturbation theory, it is also possible to distinguish between
a host of inflationary models further by comparing 
the non-Gaussianity of primordial perturbations with
the WMAP data \cite{KSpergel,Bartolo,Maldacena,Cremi2003,Rigo,Lyth,Byrnes}. 
In particular it is expected that the PLANCK 
satellite \cite{PLANCK} will bring us more precise data of non-Gaussianities
within a few years.

The amount of non-Gaussianity can be quantified by evaluating 
the bispectrum of curvature perturbations ${\cal R}$, as 
\begin{equation}
\langle {\cal R} ({\bm k}_1) {\cal R} ({\bm k}_2)
{\cal R} ({\bm k}_3) \rangle =(2\pi)^3 \delta^{(3)}
({\bm k}_1+{\bm k}_2+{\bm k}_3)
B_{\cal R} (k_1, k_2, k_3)\,,
\end{equation}
where ${\cal R} ({\bm k})$ is a Fourier component
of ${\cal R}$ with a wave number ${\bm k}$.
Conventionally the bispectrum $B_{\cal R}$ translates into 
a non-linear parameter $f_{\rm NL}$ in order to confront 
theoretically predicted non-Gaussianities 
with observations \cite{KSpergel,WMAP7}.
There are two different shapes of the primordial 
bispectrum: (1) ``local'' type \cite{Gangui,Verde,KSpergel}, 
and (2) ``equilateral'' type \cite{Cremi}.
The first one arises from a local, point-like non-Gaussianity 
given by ${\cal R} ({\bm x})={\cal R}_{\rm L}({\bm x})+(3/5)
f_{\rm NL}^{\rm local}\,{\cal R}_{\rm L}^2({\bm x})$, 
where ${\cal R}_{\rm L}$ is a linear Gaussian 
perturbation.
In single-field inflation models driven by a slowly varying potential, 
the predicted value of $f_{\rm NL}^{\rm local}$ is of the 
order of slow-roll parameters, 
i.e. $|f_{\rm NL}^{\rm local}| \ll 1$ \cite{Salopek,Gangui,Bartolo,Maldacena,Cremi2003,Rigo}.
This small amount of non-Gaussianity is consistent with 
the WMAP 7-year bound: $-10<f_{\rm NL}^{\rm local}<74$
(95 \% CL) \cite{WMAP7}.

In the equilateral shape of non-Gaussianities, 
the momentum dependence of the function $B_{\cal R} (k_1, k_2, k_3)$
arising from non-canonical kinetic term models (``k-inflation'' \cite{kinf}) 
can be approximated in a suitable form \cite{Cremi}. 
In k-inflation models, including  Dirac-Born-Infeld (DBI) 
inflation \cite{DBIinf} and (dilatonic) ghost condensate \cite{ghost}, 
one can realize the non-linear parameter 
$f_{\rm NL}^{\rm equil}$ larger than the order of unity
provided that the propagation speed $c_s$ of scalar perturbations
is much smaller than 1 
(in unit of speed of light) \cite{Gruzinov,Seery,Chen,DBIsky,Chen2,Cheung}.
The WMAP 7-year bound is 
$-214<f_{\rm NL}^{\rm equil}<266$ (95 \% CL) \cite{WMAP7}, 
but it is expected that the PLANCK satellite can reduce this limit by 
one order of magnitude. This will provide us 
an opportunity to
distinguish k-inflation from standard 
slow-roll inflation.
We also note that it is possible to give rise to large 
non-Gaussianities in multiple scalar-field 
models \cite{Linde97,Uzan,Seery2,Arroja,Langlois,multi1,multi2,multi3,Gao,Rodr}, 
curvaton models \cite{curvaton1,curvaton2,curvaton3,curvaton4,curvaton5}, 
modulated reheating models \cite{modulated,Suyama}, 
models having a preheating stage after inflation \cite{preheating}, 
and models with a temporal non-slow roll stage \cite{Taka}
(see also Refs.~\cite{otherpapers}).

In this paper we shall compute the bispectrum $B_{\cal R}$
and the equilateral-type non-linear 
parameter $f_{\rm NL}^{\rm equil}$ for general single-field 
inflation models described by the action (\ref{action}) below.
In low-energy effective string theory there is a scalar field $\phi$
called dilaton coupled to the Ricci scalar $R$ with 
a form $F(\phi)R$ \cite{Gas}.
A field coupling of the form $\xi (\phi) {\cal G}$, where
${\cal G}$ is the Gauss-Bonnet  term, also arises
as a higher-order string correction to the low-energy effective 
string action \cite{GBterm}.  Furthermore the higher-order string correction contains
a non-canonical kinetic term like $(\partial \phi)^4$ as well as
a field self-interaction of the form $g(\phi)(\partial \phi)^2 \square \phi$
in the action. For constant $g (\phi)$ the latter is linked to the 
Lagrangian of a covariant Galileon field that respects the Galilean symmetry 
$\partial_{\mu} \phi \to \partial_{\mu} \phi+b_{\mu}$
in the Minkowski space-time \cite{Nicolis,Deffayet}.
The cosmological dynamics in the presence of the Galileon-type interaction 
$g(\phi)(\partial \phi)^2 \square \phi$ have been extensively studied
recently in the context of inflation \cite{Galileoninf,Mizuno} and dark 
energy \cite{Galileondark}.
We accommodate non-linear field derivative terms
as the Lagrangian of the form $P(\phi, X)-G(\phi, X) \square \phi$, 
where $P$ and $G$ are functions in terms of $\phi$ and 
$X=-(\partial \phi)^2/2$. 
Note that non-Gaussianities in the models described by 
the Lagrangian $F(\phi)R+P(\phi, X)$ without
the terms $\xi (\phi) {\cal G}$ and $G(\phi, X) \square \phi$ 
were recently studied in Ref.~\cite{Qiu}.

Our action (\ref{action}) can be also viewed as describing a kind of modified gravitational 
theories \cite{DeFelice}. In fact, this covers the so-called scalar-tensor theories \cite{Fujii}
such as Brans-Dicke (BD) theory \cite{BDtheory}. 
Since metric $f(R)$ gravity 
is equivalent to BD theory with the BD parameter 
$\omega_{\rm BD}=0$ \cite{Ohanlon},
our results can be applied also to the Starobinsky's 
inflation model $f(R)=R+\alpha R^2$ \cite{Sta80}.
Since the field propagation speed $c_s$ is unity in BD theory, 
we will show that $|f_{\rm NL}^{\rm equil}| \ll 1$
as in conventional slow-roll inflation.
On the other hand, the terms $\xi (\phi) {\cal G}$ and 
$G(\phi, X) \square \phi$ as well as $P(\phi, X)$ 
can give rise to $c_s$ much smaller than 1.
In such theories it is possible to realize large non-Gaussianities
detectable in future high-precision observations. 

This paper is organized as follows.
In Sec.~\ref{modelsec} the field equations of motion 
are derived for the action (\ref{action})
on the flat Friedmann-Lema\^{i}tre-Robertson-Walker
(FLRW) background.
In Sec.~\ref{secondorder} we obtain the second-order 
action for the curvature perturbation ${\cal R}$ 
and present the solution of its mode function
at linear level on the de Sitter background.  
In Sec.~\ref{thirdlag} the third-order perturbed action 
is derived in order to compute
the three-point correlation function of ${\cal R}$
in the interacting Hamiltonian picture.
We also present a general analytic formula of $f_{\rm NL}^{\rm equil}$
valid in the quasi de Sitter background.
In Sec.~\ref{expansion} we provide a simpler expression of 
$f_{\rm NL}^{\rm equil}$ under the expansion of ``slow-variation''
parameters. This approximate formula is enough to estimate
the amount of non-Gaussianities in practical purpose.
In Sec.~\ref{cmodels} we apply our general results to a number of 
concrete models of inflation: (1) k-inflation, (2) generalized Galileon 
model, and (3) Brans-Dicke theory.
Sec.~\ref{conclusions} is devoted to conclusions.
In Appendix we present the detailed procedure
to derive and manipulate the third-order perturbed action.

\section{The model and background equations}
\label{modelsec} 

We start with the following action 
\begin{equation}
S=\int d^{4}x\sqrt{-g}\left[\frac{M_{{\rm pl}}^{2}}{2}F(\phi)R
+P(\phi,X)-\xi(\phi){\cal G}-G(\phi,X)\square\phi\right]\,,
\label{action}
\end{equation}
where $g$ is the determinant of the space-time metric $g_{\mu\nu}$,
$M_{{\rm pl}}=(8\pi G)^{-1/2}$ is the reduced Planck mass 
($G$ is gravitational constant), and $\phi$ is a scalar field with
a kinetic term $X=-(1/2)g^{\mu\nu}\partial_{\mu}\phi\partial_{\nu}\phi$.
The functions $F(\phi)$ and $\xi(\phi)$ depend on $\phi$ only, 
whereas $P(\phi,X)$ and $G(\phi,X)$ are functions of both $\phi$ and $X$.
The field $\phi$ couples to the Ricci scalar $R$ as well as 
the Gauss-Bonnet term ${\cal G}$ defined by 
\begin{equation}
{\cal G} \equiv R^{2}-4R_{\alpha\beta}R^{\alpha\beta}
+R_{\alpha\beta\gamma\delta}R^{\alpha\beta\gamma\delta}\,,\label{gb}
\end{equation}
where $R_{\alpha\beta}$ is the Ricci scalar and 
$R_{\alpha\beta\gamma\delta}$
is the Riemann tensor for the metric $g_{\mu\nu}$.
As we mentioned in the Introduction, the action (\ref{action}) covers
a wide variety of single-field inflationary models.

We consider the flat FLRW space-time with a scale factor $a(t)$, 
where $t$ is the cosmic time. Then the background equations are given by 
\begin{eqnarray}
E_{1} & \equiv & 
3\Mpl^{2}FH^2+3\Mpl^{2} H \dot{F}
+P-2XP_{,X}-24H^{3}\dot{\xi}-6H \dot{\phi} XG_{,X}
+2X G_{,\phi}=0\,,\\
\label{E2eq}
E_{2} & \equiv & 
3\Mpl^{2}FH^2+2\Mpl^{2} H \dot{F}+2\Mpl^2 F \dot{H}
+\Mpl^2 \ddot{F}+P-16H^{3}\dot{\xi}-16H \dot{H} \dot{\xi}
-8H^{2}\ddot{\xi}-G_{,X} \dot{\phi}^2 \ddot{\phi}
-G_{,\phi} \dot{\phi}^2=0\,,\\
E_{3} & \equiv & 
( P_{,X}+2XP_{,XX}+6H \dot{\phi} G_{,X} +6H \dot{\phi} XG_{,XX}
-2X G_{,\phi X}-2G_{,\phi} ) \ddot{\phi} \nonumber \\
& &{} +( 3H P_{,X}+\dot{\phi} P_{,\phi X}+9H^2 \dot{\phi} G_{,X}
+3\dot{H} \dot{\phi}G_{,X}+3H \dot{\phi}^2 G_{,\phi X}
-6HG_{,\phi}-G_{,\phi \phi} \dot{\phi}) \dot{\phi} \nonumber \\
& &{} -P_{,\phi}-6\Mpl^2 H^2 F_{,\phi}
-3\Mpl^2 \dot{H} F_{,\phi}
+24 H^4 \xi_{,\phi}+24H^2 \dot{H} \xi_{,\phi}=0\,,
\end{eqnarray}
where $H\equiv\dot{a}/a$ is the Hubble parameter, and a dot represents
a derivative with respect to $t$. These equations are
not independent because of the Bianchi identities, i.e. 
$\dot{\phi}E_{3}+\dot{E}_{1}+3H(E_{1}-E_{2})=0$. 
In Eq.~(\ref{E2eq}) the term $G_{,X} \dot{\phi}^2 \ddot{\phi}$ vanishes
for the theories with $G_{,X}=0$, in which case the Lagrangian $G(\phi) \square \phi$
can be regarded as a part of the Lagrangian $P(\phi, X)$.
If $G_{,X} \neq 0$, however, the term $G(\phi,X)\square\phi$ 
should be treated separately from the term $P(\phi, X)$.

In order to derive the slowly varying parameter $-\dot{H}/H^2$ 
we consider the combination 
$(E_2-E_1)/(M_{\rm Pl}^2 H^2 F)=0$, which gives
\begin{equation}
\epsilon \equiv -\frac{\dot{H}}{H^2}=-\frac{\dot{F}}{2HF}+\frac{\ddot{F}}{2H^2F}
+\frac{XP_{,X}}{\Mpl^2 H^2 F}+\frac{4H \dot{\xi}}{\Mpl^2 F}
-\frac{8 \dot{H} \dot{\xi}}{\Mpl^2 HF}-\frac{4\ddot{\xi}}{\Mpl^2 F}
+\frac{3\dot{\phi} X G_{,X}}{\Mpl^2 HF}
-\frac{\ddot{\phi}XG_{,X}}{\Mpl^2 H^2 F}
-\frac{2XG_{,\phi}}{\Mpl^2 H^2 F}\,.
\label{dotHeq}
\end{equation}
During inflation the Hubble parameter changes slowly, so that the 
condition $\epsilon \ll 1$ is satisfied.
Hence, in general, we require that each term on the r.h.s.\ of Eq.~(\ref{dotHeq})
is much smaller than unity.
In Sec.~\ref{expansion} we shall use this property to obtain 
a simple expression for the equilateral non-linear parameter $f_{\rm NL}^{{\rm equil}}$.
For later convenience we introduce the following ``slow-variation'' parameters
\begin{equation}
\delta_F \equiv \frac{\dot{F}}{HF}\,,\qquad
\delta_{\xi} \equiv \frac{H\dot{\xi}}{\Mpl^2 F}\,,\qquad
\delta_{GX} \equiv \frac{\dot{\phi} X G_{,X}}{\Mpl^2 H F}\,.
\label{slowvariation}
\end{equation}
%

\section{Second-order action and linear perturbations}
\label{secondorder} 

In order to compute primordial scalar non-Gaussianities 
we need to expand the action (\ref{action}) up to third 
order in the perturbations, taking into account, up to a gauge choice, 
both the perturbations in the scalar field, $\delta \phi$, 
and in the scalar modes of 
the metric. The interacting Hamiltonian of perturbations 
follows from the third-order Lagrangian, by which the 
three-point correlation function of curvature perturbations
${\cal R}$ can be evaluated in the framework of quantum 
field theory \cite{Maldacena}.
In order to calculate the vacuum expectation value of
the correlation function, the mode function of ${\cal R}$
should be known in the quasi de Sitter background.
Once the mode function is obtained at linear level, 
one can derive the power spectrum of ${\cal R}$
generated during inflation.
The linear perturbation equation for ${\cal R}$ is 
known by the second-order perturbed action.

For the derivation of the action expanded up to
third order in the perturbations, it is convenient to work in the ADM 
formalism \cite{ADM} with the line element
\begin{equation}
ds^{2}=-N^{2}dt^{2}+h_{ij}(N^{i}dt+dx^{i})
(N^{j}dt+dx^{j})\,,
\label{metric2}
\end{equation}
where $N$ and $N^i$ are the lapse and shift functions, respectively.
Here we consider only scalar metric perturbations 
about the flat FLRW background. 
In doing so we expand the lapse $N$ and the shift vector
$N^{i}$, as $N=1+\alpha$ and $N_{i}=\partial_{i}\psi$, respectively ($i=1, 2, 3$).
In fact, in the ADM formalism, the lapse $N$ and the shift $N^i$
are Lagrange multipliers, so that it is sufficient to know $N$ and $N^i$
up to first order. This is because the third-order and the second-order 
terms in $N, N^i$ multiply the constraint equations at zero-th order
and at first order, respectively, so that their contributions 
vanish \cite{Maldacena,Koyama}. 
We choose the uniform-field gauge 
with $\delta\phi=0$, which fixes the time-component
of a gauge-transformation vector $\xi^{\mu}$. We gauge away a field
$E$ that appears as a form $E_{,ij}$ inside $h_{ij}$, by fixing
the spatial part of $\xi^{\mu}$. Then the three-dimensional metric
can be written as $h_{ij}=a^{2}(t)e^{2{\cal R}}\delta_{ij}$. 
This results in the following metric
\begin{equation}
ds^{2}=-\left[ (1+\alpha)^{2}-a^{-2}(t) e^{-2{\cal R}}(\partial\psi)^{2}
\right]\, dt^{2}+2\partial_{i}\psi\, dt\, dx^{i}+a^2(t) e^{2{\cal R}}
(dx^{2}+dy^{2}+dz^{2})\,,
\label{eq:metrica}
\end{equation}
where $(\partial\psi)^{2}=(\partial\psi_i)\,(\partial\psi_i)\equiv(\partial_{x}\psi)^{2}
+(\partial_{y}\psi)^{2}+(\partial_{z}\psi)^{2}$. 
Here and in the following, same lower latin indices are summed, 
unless otherwise specified.
At linear level the metric (\ref{eq:metrica}) reduces to the standard one used 
in linear perturbation theory, that is \cite{Bardeen} 
\begin{equation}
ds^{2}=-(1+2\alpha)\, dt^{2}+2\partial_{i}\psi\, dt\, dx^{i}
+a^2(t)\, (1+2{\cal R})\,(dx^{2}+dy^{2}+dz^{2})\,.
\end{equation}

Expanding the action (\ref{action}) up to second order, one finds
\begin{equation}
\label{eq:lag2a}
S_2=\int dt\,d^3x\,a^3
\left[ -3w_1 \dot{\cal R}^2 
+\frac{2w_1}{a^2} \dot{\cal R} \partial^2\psi
-\frac{w_2}{a^2} \alpha \partial^2\psi
-\frac{2w_1}{a^2} \alpha \partial^2{\cal R}
+3w_2\,\alpha\,\dot{\cal{R}}
+\frac13 w_3 \alpha^2
+\frac{w_4}{a^2}\,\partial_i{\cal R}\,\partial_i{\cal R}\right]\,,
\end{equation}
where
\begin{eqnarray}
w_{1} & \equiv & \Mpl^{2}\, F-8H\,\dot{\xi}\,,\\
w_{2} & \equiv & \Mpl^{2}(2HF+\dot{F})
-2\dot{\phi} XG_{,X}-24H^{2}\dot{\xi}\,,\\
w_3 &\equiv& 
-9\Mpl^2 F{H}^{2}-9\Mpl^2 H \dot{F}
+3(XP_{,X}+2X^2 P_{,XX})+144H^3 \dot{\xi}
\nonumber \\
& &{}+18 H\dot{\phi} (2XG_{,X}+X^2 G_{,XX})
-6(X G_{,\phi}+X^2 G_{,\phi X})\,,\\
w_4&\equiv& \Mpl^{2}F-8\ddot{\xi}\,.
\end{eqnarray}
In the action (\ref{eq:lag2a}), both the coefficients of the terms 
$\alpha {\cal R}$ and ${\cal R}^2$ vanish by using the background equations of motion. 
This is related to the fact that the field ${\cal R}$ does not have an explicit mass term.
Furthermore, in (\ref{eq:lag2a}), the term quadratic in $\psi$ vanishes after integrations by parts.
The equations of motion for $\psi$ and $\alpha$, derived from (\ref{eq:lag2a}),
lead to the following two constraints 
\begin{eqnarray}
\alpha &=& L_{1} \dot{{\cal R}}\,,\label{eq:constroA}\\
\frac{1}{a^2}\,\partial^2\psi &=& \frac{2w_3}{3w_2}\,\alpha
+3\dot{\cal R}-\frac{2w_1}{w_2}\frac1{a^2}
\partial^2{\cal R}\,,
\label{eq:constroB}
\end{eqnarray}
where 
\begin{equation}
L_1 \equiv \frac{2w_1}{w_2}=\frac{2(\Mpl^2 F-8 H\dot{\xi})}
{\Mpl^2 (2HF+\dot{F})-2\dot{\phi} XG_{,X}-24H^{2}\dot{\xi}}\,.
\end{equation}
In k-inflation with $F=1$, $\xi=0$, and $G=0$
one has $L_1=1/H$.
In general cases, expansion in terms of the slow-variation parameters defined in 
Eq.~(\ref{slowvariation}) gives 
\begin{equation}
L_1 =\frac{1}{H} \left[1-\frac12 \delta_F+4\delta_{\xi}+
\delta_{GX}+{\cal O} (\epsilon^2) \right]\,.
\label{L1expansion}
\end{equation}

Plugging the relation $\alpha=(2w_1/w_2)  \dot{\cal R}$ 
into Eq.~(\ref{eq:lag2a}) and integrating the term 
$\dot{\cal R}\partial^2{\cal R}$ by parts as 
\begin{equation}
c(t)\dot{\cal R}\partial^2{\cal R}=\dot c\,(\partial{\cal R})^2/2
+{\rm total~derivatives}\,,
\end{equation}
one finds
\begin{equation}
S_{2}=\int dt\, d^{3}x\, a^{3}Q\left[\dot{{\cal R}}^{2}
-\frac{c_{s}^{2}}{a^{2}}\,(\partial{\cal R})^{2}\right]\,,
\label{eq:az2}
\end{equation}
where
\begin{eqnarray}
Q & \equiv& \frac{w_1 (4w_1w_3+9w_2^2)}{3w_2^2}\,,\label{eq:defQ}\\
c_s^2 &\equiv& 
\frac{3(2 w_1^{2} w_2H-w_2^2 w_4+4 w_1 \dot{w}_1w_2
-2w_1^{2}\dot{w}_2)}{w_1(4w_1w_3+9w_2^2)}\,.
\label{eq:defc2s}
\end{eqnarray}
More explicit expressions for $Q$ and $c_s^2$ are given in Appendix \ref{app:actio3}. 
In order to avoid the appearance of ghosts and Laplacian instabilities
we require that $Q>0$ and $c_{s}^{2}>0$, respectively.
Using Eqs.~(\ref{eq:constroA}) and (\ref{eq:defQ}), we can rewrite 
Eq.~(\ref{eq:constroB}) as
\begin{equation}
\psi=-L_{1} {\cal R}+\chi\,,\qquad{\rm where}
\qquad\partial^{2}\chi=a^{2}\, \frac{Q}{w_1}\,\dot{{\cal R}}\,.
\label{eq:allin}
\end{equation}

We introduce a parameter $\epsilon_s$ defined by 
\begin{equation}
\epsilon_s \equiv \frac{Qc_s^2}{\Mpl^2 F}
=\frac{2 w_1^{2} w_2H-w_2^2 w_4+4 w_1 \dot{w}_1w_2
-2w_1^{2}\dot{w}_2}{\Mpl^2 F w_2^2}\,.
\label{epsilonsdef}
\end{equation}
In k-inflation with $F=1$, $\xi=0$, and $G=0$
this reduces to the slow-roll parameter 
$\epsilon=-\dot{H}/H^{2}$.
In general, one can expand $\epsilon_s$ in terms of 
slow-variation parameters, as
\begin{equation}
\epsilon_s=\epsilon+\frac12 \delta_F+\delta_{GX}
-4\delta_{\xi}+{\cal O} (\epsilon^2)\,.
\label{epsilons}
\end{equation}

The equation of motion for ${\cal R}$ follows
by varying the Lagrangian ${\cal L}_{2}=
a^{3}Q [\dot{{\cal R}}^{2}-(c_{s}^{2}/a^{2})\,(\partial{\cal R})^{2}]$
in terms of ${\cal R}$. We define 
\begin{equation}
\frac{\delta{\cal L}_{2}}{\delta{\cal R}}\biggr|_{1} 
\equiv -2 \left[\frac{d}{dt}(a^3 Q \dot{\cal R})
-aQ c_s^2 \partial^2 {\cal R} \right]\,.
\label{linearR}
\end{equation}
Then the curvature perturbation obeys
the equation 
$\delta {\cal L}_2/\delta {\cal R}|_1=0$ 
at linear level, i.e.
\begin{equation}
\frac{d}{dt}(a^3 Q \dot{\cal R})
-aQ c_s^2 \partial^2 {\cal R}=0\,.
\label{linearR2}
\end{equation}
We write ${\cal R}$ in Fourier space, as 
\begin{equation}
{\cal R}(\tau,{\bm{x}})=\frac{1}{(2\pi)^{3}}\int d^{3}{\bm{k}}{\cal R}(\tau,{\bm{k}})e^{i{\bm{k}}\cdot{\bm{x}}}\,,\qquad{\cal R}(\tau,{\bm{k}})=u(\tau,{\bm{k}})a({\bm{k}})+u^{*}(\tau,{-\bm{k}})a^{\dagger}(-{\bm{k}})\,,
\label{RFourier}
\end{equation}
where $a({\bm{k}})$ and $a^{\dagger}({\bm{k}})$ are the annihilation
and creation operators, respectively, satisfying the commutation relations
\begin{equation}
\left[a({\bm{k}}_{1}),a^{\dagger}({\bm{k}}_{2})\right]
=(2\pi)^{3}\delta^{(3)}({\bm{k}}_{1}-{\bm{k}}_{2})\,,
\qquad\left[a({\bm{k}}_{1}),a({\bm{k}}_{2})\right]
=\left[a^{\dagger}({\bm{k}}_{1}),a^{\dagger}({\bm{k}}_{2})\right]=0\,.
\end{equation}
Note that $\tau \equiv \int a^{-1}dt$ is the conformal time, which is given
by $\tau=-1/(aH)$ in the de Sitter background.
The asymptotic past and future correspond to $\tau \to -\infty$ 
and $\tau \to -0$, respectively.

Let us derive the solution of ${\cal R}$ during inflation
at linear order. The equation for the Fourier mode $u$ is given by 
\begin{equation}
\ddot{u}+\frac{(a^{3}Q)^{\cdot}}{a^{3}Q}\dot{u}
+c_{s}^{2}\frac{k^{2}}{a^{2}}u=0\,.
\label{ueq}
\end{equation}
In the large-scale limit ($k \to 0$) the solution to this equation is 
$u=c_1+c_2 \int (a^3 Q)^{-1}\,dt$, where $c_1$ and $c_2$ are 
integration constants. Provided that the variable $Q$ changes slowly
in time, $u$ approaches a constant after the perturbations 
exit the Hubble radius ($c_s k \lesssim aH$).
Introducing a field $v=zu$ with $z=a\sqrt{2Q}$
the kinetic term in the second-order action (\ref{eq:az2})
can be rewritten as $\int d\tau d^3 x\,v'^2/2$, 
where a  prime represents a derivative with respect to $\tau$. 
In other words, $v$ is the canonical field that should be quantized.
Equation (\ref{ueq}) can be written as 
\begin{equation}
v''+\left(c_{s}^{2}k^{2}-\frac{z''}{z}\right)v=0\,.
\label{veq}
\end{equation}
In the de Sitter background with a slow variation of 
the quantity $Q$, we can approximate $z''/z \simeq 2/\tau^{2}$. 
In the asymptotic past ($k\tau\to-\infty$) we choose 
the Bunch-Davis vacuum characterized by the mode function 
$v=e^{-ic_{s}k\tau}/\sqrt{2c_{s}k}$.
Then the solution of Eq.~(\ref{veq}) is given by 
\begin{equation}
u (\tau, k)=\frac{i\,H\, e^{-ic_{s}k\tau}}
{2(c_{s}k)^{3/2}\sqrt{Q}}\,(1+ic_{s}k\tau)\,.
\label{usol}
\end{equation}
The deviation from the exact de Sitter background gives 
rise to a small modification to the solution (\ref{usol}), 
but this difference appears as a next-order correction 
to the power spectrum and to the non-Gaussianity 
parameter \cite{Chen}.
 
The two-point correlation function, some time after the 
Hubble radius crossing, is given by the vacuum expectation value
$\langle 0| {\cal R} (\tau, {\bm k}_1) {\cal R} (\tau,{\bm k}_2) | 0 \rangle$
at $\tau \approx 0$.
We define the power spectrum ${\cal P}_{\cal R} (k_1)$, as 
$\langle 0| {\cal R} (0,{\bm k}_1) {\cal R} (0,{\bm k}_2) | 0 \rangle
=(2\pi^2/k_1^3){\cal P}_{\cal R} (k_1)\,
(2\pi)^3 \delta^{(3)} ({\bm k}_1+{\bm k}_2)$.
Using the solution (\ref{usol}), we obtain
\begin{equation}
{\cal P}_{\cal R}=\frac{H^2}{8\pi^2 Q c_s^3}
=\frac{H^2}{8\pi^2 \Mpl^2 F \epsilon_s c_s}\,,
\label{scalarpower}
\end{equation}
where we have used $\epsilon_s$ defined in Eq.~(\ref{epsilonsdef}).
Since the curvature perturbation soon approaches a constant 
for $c_s k <aH$, we just need to evaluate
the power spectrum (\ref{scalarpower}) at
$c_s k=aH$ during inflation \cite{Garriga}.
The spectral index of ${\cal R}$ is given by 
\begin{eqnarray}
n_{\cal R}-1 \equiv \frac{d \ln {\cal P}_{\cal R}}
{d \ln k}\bigg|_{c_sk=aH}
&=& -2\epsilon-\delta_F-\eta_s-s \nonumber \\
&=& -2\epsilon_s-\eta_s-s-8\delta_{\xi}+2\delta_{GX}\,,
\label{nR}
\end{eqnarray}
where 
\begin{equation}
\eta_s \equiv \frac{\dot{\epsilon}_s}{H \epsilon_s}\,,\qquad
s \equiv \frac{\dot{c}_s}{H c_s}\,.
\label{etas}
\end{equation}
Here we assumed that both $H$ and $c_s$ slowly vary, 
such that $d \ln k$ at $c_s k=aH$ is 
approximated as $d \ln k=d \ln a=H dt$.
In the second equality of Eq.~(\ref{nR}) we used Eq.~(\ref{epsilons})
to convert $\epsilon$ to $\epsilon_s$ at linear order.

Let us also derive the spectrum of tensor perturbations generated during inflation.
In addition to the three-dimensional metric $a^2(t)e^{2{\cal R}} \delta_{ij}$
coming from the scalar part, we consider the intrinsic tensor perturbation
$h_{ij}^{(T)}$. 
In terms of the two polarization tensors $e_{ij}^{+}$ and $e_{ij}^{\times}$
we can write $h_{ij}^{(T)}=h_{+}e_{ij}^{+}+h_{\times} e_{ij}^{\times}$, 
where both the symmetric tensors $e_{ij}$ are transverse and traceless. 
We also impose the normalization condition, 
$e_{ij} (\bm{k})\,e_{ij} (-\bm{k})^{*}=2$, for each polarization, 
whereas $e^{+}_{ij} (\bm{k})\,e^{\times}_{ij}(-\bm{k})^{*}=0$.
The second-order action for the tensor modes is given by 
\begin{equation}
S_{T}=\sum_{\lambda}\int dt\, d^{3}x\, a^{3} Q_{T}\left[ \dot{h}_{\lambda}^{2}
-\frac{c_{T}^2}{a^2} (\partial h_{\lambda})^2 \right]\,,
\label{ST}
\end{equation}
where $\lambda=+, \times$, and 
\begin{eqnarray}
& &Q_{T}=\frac{1}{4}\, w_{1}=\frac{1}{4}\, M_{{\rm pl}}^{2}F\,
(1-8\delta_{\xi})\,,
\label{QT} \\ 
& &c_{T}^{2}=\frac{w_{4}}{w_{1}}=\frac{M_{{\rm pl}}^{2}F
-8\ddot{\xi}}{M_{{\rm pl}}^{2}F-8H\dot{\xi}}
=1+8\delta_{\xi}+{\cal O} (\epsilon^2)\,.
\label{cT}
\end{eqnarray}

A canonical field associated with $h_{\lambda}$ corresponds to 
$v_T=z_T h_{\lambda}$ and $z_T=a\sqrt{2Q_T}$. 
Following the same procedure as before, 
the solution to $h_{\lambda}$ recovering the Bunch-Davis 
vacuum in the asymptotic past is 
\begin{equation}
h_{\lambda}=\frac{i\,H\, e^{-ic_{T}k\tau}}
{2(c_{T}k)^{3/2}\sqrt{Q_T}}\,(1+ic_{T}k\tau)\,,
\label{hsol}
\end{equation}
which approaches $h_{\lambda} \to i H/[2(c_Tk)^{3/2} \sqrt{Q_T}]$
after the Hubble radius crossing.
According to the chosen normalization for the tensors 
$e^{\lambda}_{ij}$, the two-point correlation function leads to
a tensor power spectrum, whose expression is given by 
${\cal P}_T=4 \cdot k^3 |h_{\lambda}|^2/(2\pi^2)$, i.e.
\begin{equation}
{\cal P}_T=\frac{H^2}{2\pi^2 Q_T c_T^3}
\simeq \frac{2H^2}{\pi^2 \Mpl^2 F}\,,
\label{tensorpower}
\end{equation}
where in the last approximate equality we have 
used Eqs.~(\ref{QT}) and (\ref{cT}) at leading order.
The spectral index of ${\cal P}_T$ is
\begin{eqnarray}
n_T\equiv \frac{d \ln {\cal P}_T}
{d \ln k}\bigg|_{c_T k=aH}
&=& -2\epsilon-\delta_F \nonumber \\
&=& -2\epsilon_s-8\delta_{\xi}+2\delta_{GX}\,,
\label{nT}
\end{eqnarray}
which is valid up to first order of slow-variation 
parameters.

For those times before the end of inflation ($\epsilon \ll 1$)
when both ${\cal P}_{\cal R}$ and ${\cal P}_{T}$ remain 
approximately constants, 
we can estimate the tensor-to-scalar ratio as
\begin{equation}
r \equiv \frac{{\cal P}_T}{{\cal P}_{\cal R}}
=4\frac{Q c_s^3}{Q_T c_T^3}
\simeq 16 c_s \epsilon_s\,,
\label{rfi}
\end{equation}
where in the last approximate equality the terms at second order 
of slow-variation parameters are not taken into account.
Using Eqs.~(\ref{nT}) and (\ref{rfi}), we obtain the consistency relation 
\begin{equation}
r =8c_s \left( -n_T-8\delta_{\xi}+2\delta_{GX}
\right)\,.
\end{equation}

The Gauss-Bonnet term, as well as the Galileon term, modifies 
the consistency relation $r=-8c_s n_T$ valid
in k-inflation \cite{Garriga}.

\section{Third-order action and primordial non-Gaussianities}
\label{thirdlag}

In order to evaluate the three-point correlation function 
of curvature perturbations, we need to expand the action~(\ref{action}) 
up to third order in the perturbation fields.
From the computational point of view, the fact that we just need to 
expand the lapse and shift only up to first order 
(as already stated in Sec.~\ref{secondorder}) is very useful for
deriving the action at cubic order.

After many integrations by parts, the third-order action 
following from (\ref{action}) can be written as
\begin{eqnarray}
S_{3} & = & \int dt\, d^{3}x\, a^{3}\,\{a_{1}\,\alpha^{3}+\alpha^{2}\,(a_{2}\,{\cal R}+a_{3}\,\dot{{\cal R}}+a_{4}\,\partial^{2}{\cal R}/a^{2}+a_{5}\partial^{2}\psi/a^{2})\nonumber \\
 &  & {}+\alpha\,[a_{6}\,\partial_{i}{\cal R}\partial_{i}\psi/a^{2}+a_{7}\,\dot{{\cal R}}{\cal R}+a_{8}\,\dot{{\cal R}}\partial^{2}{\cal R}/a^{2}+a_{9}\,(\partial_{i}\partial_{j}\psi\partial_{i}\partial_{j}\psi-\partial^{2}\psi\partial^{2}\psi)/a^{4}
\nonumber \\
& & {}+a_{10}(\partial_{i}\partial_{j}\psi\partial_{i}\partial_{j}{\cal R}-\partial^{2}\psi\partial^{2}{\cal R})/a^{4}
+a_{11}\,{\cal R}\,\partial^{2}\psi/a^{2}+a_{12}\,\dot{{\cal R}}\,\partial^{2}\psi/a^{2}+a_{13}\,{\cal R}\,\partial^{2}{\cal R}/a^{2}+a_{14}\,(\partial{\cal R})^{2}/a^{2}+a_{15}\dot{{\cal R}}^{2}]\nonumber \\
 &  & {}+b_{1}\,\dot{{\cal R}}^{3}+b_{2}\,{\cal R}\,(\partial{\cal R})^{2}/a^{2}+b_{3}\dot{{\cal R}}^{2}\,{\cal R}+c_{1}\,\dot{{\cal R}}\partial_{i}{\cal R}\partial_{i}\psi/a^{2}+c_{2}\dot{{\cal R}}^{2}\partial^{2}\psi/a^{2}+c_{3}\dot{{\cal R}}\,{\cal R}\,\partial^{2}\psi/a^{2}\nonumber \\
 &  & {}+(d_{1}\dot{{\cal R}}+d_{2}{\cal R})\,(\partial_{i}\partial_{j}\psi
\partial_{i}\partial_{j}\psi-\partial^{2}\psi\partial^{2}\psi)/a^{4}+d_{3}\partial_{i}
{\cal R}\partial_{i}\psi\,\partial^{2}\psi/a^{4}\}\,,
\label{eq:S3}
\end{eqnarray}
where
\begin{eqnarray}
a_1 &=& 3M_{\rm pl}^{2}F{H}^{2}+3 M_{\rm pl}^{2}\dot F H
-XP_{,X}-4X^2 P_{,XX}-4X^3 P_{,XXX}/3-80 H^{3}\dot\xi \nonumber \\
& &{} -2H \dot{\phi} (10XG_{,X}+11X^2 G_{,XX}+2X^3 G_{,XXX})
+2XG_{,\phi}+14X^2 G_{,\phi X}/3+4X^3 G_{,\phi XX}/3 \,,\\
a_2 &=& w_3=-9\Mpl^2 F{H}^{2}-9\Mpl^2 H \dot{F}
+3(XP_{,X}+2X^2 P_{,XX})+144H^3 \dot{\xi}
\nonumber \\
& &~~~~~~~\,+18 H\dot{\phi} (2XG_{,X}+X^2 G_{,XX})
-6(X G_{,\phi}+X^2 G_{,\phi X})\,,\\
a_{3} &=& -3a_5= -3 [ 2M_{\rm pl}^2 FH+M_{\rm pl}^2
\dot{F}-48 H^2 \dot{\xi} -2\dot{\phi} (2X G_{,X}+X^2 G_{,XX})]\,,\\
a_{4} & = & -16H\dot{\xi}\,,\\
a_{6} & = &-a_7/9=a_{11}=-w_2= 
-[\Mpl^{2}(2HF+\dot{F})
-2\dot{\phi} XG_{,X}-24H^{2}\dot{\xi}]\,,\\
a_{8} & = &2a_{10}=2b_1=-2c_2=-4d_1= 16\dot{\xi}\,,\\
a_{9} & = &a_{12}/4=-a_{15}/6= -(M_{{\rm Pl}}^{2}F-24H\dot{\xi})/2\,,\\
a_{13} & = &2a_{14}=2b_3/9=-c_1=-c_3=
-4d_2/3=d_3= -2w_1=-2(M_{{\rm Pl}}^{2}F-8H\dot{\xi})\,,\\
b_{2} & = &w_4= M_{{\rm Pl}}^{2}F-8\ddot{\xi}\,.
\end{eqnarray}
In Appendix \ref{app:intbp} we present more details for the derivation 
of Eq.~(\ref{eq:S3}).
Terms of the forms ${\cal R}^{3}$ and $\alpha{\cal R}^{2}$ 
vanish because of the background equations of motion,
which is similar to what happens for the terms 
${\cal R}^2$ and $\alpha{\cal R}$ in the second-order Lagrangian (\ref{eq:lag2a}). 
Another analogy between the two actions $S_2$ and $S_3$ consists of the cancellation
of the self-coupling term for the field $\psi$: that is why the cubic term in $\psi$ 
is absent in Eq.~(\ref{eq:S3}). 
In addition to the curvature perturbation ${\cal R}$
the action (\ref{eq:S3}) involves the terms $\alpha$ and $\psi$.
Using the constraint equation (\ref{eq:constroA}) to eliminate $\alpha$, 
the action (\ref{eq:S3}) reduces to 
\begin{eqnarray}
S_{3} & = & \int dt\, d^{3}x\, a^{3}\{A_{1}\dot{{\cal R}}^{3}+A_{2}\dot{{\cal R}}^{2}\partial^{2}{\cal R}/a^{2}+A_{3}\dot{{\cal R}}^{2}\partial^{2}\psi/a^{2}+A_{4}{\cal R}\dot{{\cal R}}^{2}+(A_{5}\dot{{\cal R}}+A_{6}{\cal R})\,(\partial_{i}\partial_{j}\psi\partial_{i}\partial_{j}\psi-\partial^{2}\psi\partial^{2}\psi)/a^{4}\nonumber \\
 &  &{} +A_{7}\dot{{\cal R}}(\partial_{i}\partial_{j}\psi\partial_{i}\partial_{j}{\cal R}-\partial^{2}\psi\partial^{2}{\cal R})/a^{4}
+A_{8}{\cal R}(\partial{\cal R})^{2}/a^{2}+A_{9}\partial_{i}{\cal R}\partial_{i}\psi\,\partial^{2}\psi/a^{4}\}\,,
\label{S3first}
\end{eqnarray}
where 
\begin{align}
A_{1} & =  b_{1}+L_{1}a_{15}+L_{1}^{2}a_{3}+L_{1}^{3}a_{1}\,,
&A_{2} & =  L_{1}\,(L_{1}a_{4}+a_{8})\,,
&A_{3} & = c_{2}+L_{1}a_{12}+L_{1}^{2}a_{5}\,,\nonumber \\
A_{4} & =  b_{3}+L_{1}a_{7}+L_{1}^{2}a_{2}\,,
&A_{5} & =  L_{1}a_{9}+d_{1}\,,
&A_{6} & =  d_{2}\,, \nonumber \\
A_{7} & =  L_{1}\, a_{10}\,,
&A_{8} & =  b_{2}+a_{13}\dot{L}_{1}/2
+L_{1}(\dot{a}_{13}+Ha_{13})/2\,,
&A_{9} & =  d_{3}\,.
\end{align}
For the derivation of Eq.~(\ref{S3first}) the term proportional to 
${\cal R}\dot{\cal R}\partial^2{\cal R}$ has been integrated by parts, as 
\begin{equation}
c(t){\cal R}\dot{\cal R}\partial^2{\cal R}=
\dot{c}(t) {\cal R}(\partial {\cal R})^2/2
-c(t)\dot{\cal R}(\partial{\cal R})^2/2
+\rm{total~derivatives}\,.
\end{equation}

The next step is to eliminate $\psi$ in the action (\ref{S3first})
by using the relation (\ref{eq:allin}).
In doing so, we introduce a new perturbation variable ${\cal X}$, 
as ${\cal X}=w_{1}\chi/a^{2}$
(in which case $\partial^{2}{\cal X}=Q\,\dot{{\cal R}}$). 
This definition is suitable in order to obtain the contributions 
proportional to the linear differential equation of ${\cal R}$, 
coming from terms which are
in the form $\partial^{2}\dot{{\cal X}}$. Substituting the relation
$\psi=-L_{1}{\cal R}+a^{2}{\cal X}/w_{1}$ into the action (\ref{S3first}),
it follows that
\begin{equation}
S_{3}=\int dt\, d^{3}x\left(a^{3}f_{1}+af_{2}
+f_{3}/a \right)\,,
\label{S3simple}
\end{equation}
where
\begin{eqnarray}
f_{1} & \equiv & \left(A_{1}+A_{3}\frac{Q}{w_{1}}-A_{5}\frac{Q^{2}}{w_{1}^{2}}\right)\dot{{\cal R}}^{3}+\left(A_{4}-A_{6}\frac{Q^{2}}{w_{1}^{2}}\right){\cal R}\dot{{\cal R}}^{2}+A_{9}\frac{Q}{w_{1}^2}\dot{{\cal R}}\partial_{i}{\cal R}\partial_{i}{\cal X}\nonumber \\
&  &{} +\frac{1}{w_{1}^{2}}\left(A_{5}\dot{{\cal R}}+A_{6}{\cal R}\right)(\partial_{i}\partial_{j}{\cal X})(\partial_{i}\partial_{j}{\cal X})\,,
\label{eq:f1expr}\\
f_{2} & \equiv & \left(A_{2}-A_{3}L_{1}+A_{5}\frac{2L_{1}Q}{w_{1}}-A_{7}\frac{Q}{w_{1}}\right)\dot{{\cal R}}^{2}\partial^{2}{\cal R}+A_{6}\frac{2L_{1}Q}{w_{1}}{\cal R}\dot{{\cal R}}\partial^{2}{\cal R}+A_{8}{\cal R}(\partial{\cal R})^{2}-A_{9}\frac{L_{1}Q}{w_{1}}\dot{{\cal R}}(\partial{\cal R})^{2} \nonumber \\
& &{}+\frac{A_{7}-2A_{5}L_{1}}{w_{1}} \dot{{\cal R}}
(\partial_{i}\partial_{j}{\cal R})(\partial_{i}\partial_{j}{\cal X})
-\frac{2A_{6}L_{1}}{w_{1}}\,{\cal R}(\partial_{i}\partial_{j}{\cal R})(\partial_{i}\partial_{j}{\cal X})-\frac{A_{9}L_{1}}{w_{1}}\,\partial^{2}{\cal R}\partial_{i}{\cal R}\partial_{i}{\cal X}\,,\label{eq:f2expr}\\
f_{3} & \equiv & \left(A_{5}L_{1}^{2}-A_{7}L_{1}\right)\dot{{\cal R}}\,[(\partial_{i}\partial_{j}{\cal R})(\partial_{i}\partial_{j}{\cal R})-(\partial^{2}{\cal R})^{2}]+A_{6}L_{1}^{2}{\cal R}\,[(\partial_{i}\partial_{j}{\cal R})(\partial_{i}\partial_{j}{\cal R})-(\partial^{2}{\cal R})^{2}]+A_{9}L_{1}^{2}(\partial{\cal R})^{2}\partial^{2}{\cal R}\,.\label{eq:f3expr}
\end{eqnarray}
In standard inflation with a canonical kinetic term 
(i.e. $F=1$, $P=X-V(\phi)$, $\xi=0$, $G=0$) one has 
$A_{1}=-M_{\rm pl}^2\epsilon/H$, $A_2=A_3=A_7=0$, 
$A_{4}=3M_{\rm pl}^2\epsilon$, $A_{5}=-M_{\rm pl}^2/(2H)$, 
$A_{6}=3M_{\rm pl}^2/2$, $A_{8}=-M_{\rm pl}^2 \epsilon$,
$A_{9}=-2M_{\rm pl}^2$, $L_{1}=1/H$, $Q=M_{\rm pl}^2 \epsilon$, and
$w_{1}=M_{\rm pl}^2$, where 
$\epsilon=\dot{\phi}^{2}/(2H^{2}M_{\rm pl}^2)$.
It then follows that 
\begin{eqnarray}
f_{1} & = & M_{\rm pl}^2 \epsilon (1-\epsilon/2 )\dot{{\cal R}}^{2}
(3{\cal R}-\dot{{\cal R}}/H)-2\epsilon\dot{{\cal R}}
\partial_{i}{\cal R}\partial_{i}{\cal X}+(3{\cal R}-\dot{{\cal R}}/H)
(\partial_{i}\partial_{j}{\cal X})(\partial_{i}\partial_{j}{\cal X})/(2M_{\rm pl}^2)\,,\\
f_{2} & = & -M_{\rm pl}^2 \epsilon ({\cal R}-2\dot{{\cal R}}/H )
(\partial{\cal R})^{2}+(M_{\rm pl}^2 \epsilon/H)
(3{\cal R}-\dot{{\cal R}}/H )\dot{{\cal R}}\partial^{2}{\cal R}
-(1/H)(3{\cal R}-\dot{\cal R}/H)(\partial_{i}\partial_{j}{\cal R})
(\partial_{i}\partial_{j}{\cal X}) \nonumber \\
& &{}+(2/H)\partial^{2}{\cal R}\partial_{i}{\cal R}\partial_{i}{\cal X}\,,\\
f_{3} & = & M_{\rm pl}^2/(2H^2) (3{\cal R}-\dot{\cal R}/H)
\left[(\partial_{i}\partial_{j}{\cal R})(\partial_{i}\partial_{j}{\cal R})
-(\partial^{2}{\cal R})^{2}\right]
-(2M_{\rm pl}^2/H^2)(\partial{\cal R})^{2}\partial^{2}{\cal R}\,,
\end{eqnarray}
which agree with Eq.~(114) in Ref.~\cite{Hael} in units where
$M_{\rm pl}^2=1$.

Although the expressions for the building blocks of the third-order action are correctly given in Eqs.~(\ref{eq:f1expr})-(\ref{eq:f3expr}), it is more convenient to perform several integrations by parts to bring each contribution into a simpler and more usable form. In Appendix \ref{app:intf3} we show that the integrand $f_3/a$ in Eq.~(\ref{S3simple}) reduces to 
\begin{equation}
f_{3}/a= (q_3/a)\,\{(\partial{\cal R})^{2}\,(\partial^{2}{\cal R})-{\cal R}\,\partial_{i}\partial_{j}[(\partial_{i}{\cal R})\,(\partial_{j}{\cal R})]\}+{\rm total~derivatives}\,,
\label{eq:f3expr2}
\end{equation}
where
\begin{equation}
q_{3}=A_{6}L_{1}^{2}-\frac{a}{3}\,\frac{d}{dt}\!
\left(\frac{A_{5}L_{1}^{2}-A_{7}L_{1}}{a}\right)+\frac{2}{3}\, A_{9}L_{1}^{2}\,.
\end{equation}
Along the same lines the term $af_2$ can be written as
(see Appendix \ref{app:intf2} for details) 
\begin{eqnarray}
af_{2} & = & -\frac{A_{7}-2A_{5}L_{1}}{2w_{1}a^{2}}
\frac{d}{dt}(a^{3}Q\,\dot{{\cal R}})\,\{(\partial{\cal R})^{2}-\partial^{-2}\partial_{i}\partial_{j}[(\partial_{i}{\cal R})(\partial_{j}{\cal R})]\}+\frac{aq_2}{2}\,\{(\partial_{i}{\cal X})\,(\partial_{i}{\cal R})\,(\partial^{2}{\cal R})-{\cal R}\,\partial_{i}\partial_{j}[(\partial_{i}{\cal R})\,(\partial_{j}{\cal X})]\}\nonumber \\
 &  & {}-2aL_1 Q {\cal R}\dot{{\cal R}}\partial^{2}{\cal R}+a\left(A_{2}-A_{3}L_{1}\right)\dot{{\cal R}}^{2}\partial^{2}{\cal R}
 +a \left[A_{8}+\frac{1}{a}\frac{d}{dt}(aL_1 Q)
 \right] {\cal R}(\partial{\cal R})^{2}+{\rm total~derivatives}\,,
\label{eq:f2expr2}
\end{eqnarray}
where
\begin{equation}
q_2=-\frac{4A_{6}L_{1}}{w_{1}}
-a^2\frac{d}{dt}\!\left(\frac{A_{7}-2A_{5}L_{1}}
{a^2w_{1}}\right)-\frac{2A_{9}L_{1}}{w_{1}}\,.
\end{equation}

Finally, the term $a^3f_1$, as shown in Appendix \ref{app:intf1}, 
is equivalent to 
\begin{eqnarray}
\hspace{-0.3cm}a^{3}f_{1} & = & a^{3}\left[A_{4}+
q_{1}(\dot{Q}+3HQ)-Q\dot{q}_{1}\right] {\cal R}\dot{{\cal R}}^{2}
-2q_{1}{\cal R}\dot{{\cal R}}\,\frac{d}{dt}(a^{3}Q\dot{{\cal R}})\nonumber \\
\hspace{-0.3cm} &  &{} +a^{3}\left(A_{1}+A_{3}\frac{Q}{w_{1}}-q_{1}\, Q\right)\dot{{\cal R}}^{3}+\frac{a^{3}}{w_{1}}\left[A_{9}\frac{Q}{w_{1}}+\frac{QA_{6}}{w_{1}}-Qw_{1}\,\frac{d}{dt}\left(\frac{A_{5}}{w_{1}^{2}}\right)+\frac{3QHA_{5}}{w_{1}}\right]\dot{{\cal R}}\partial_{i}{\cal R}\partial_{i}{\cal X}\nonumber \\
\hspace{-0.3cm} &  &{} +\frac{a^{3}}{2}\!\left[\frac{A_{6}}{w_{1}^{2}}-\frac{d}{dt}\!\left(\frac{A_{5}}{w_{1}^{2}}\right)+\frac{3HA_{5}}{w_{1}^{2}}\right]\!(\partial^{2}{\cal R})(\partial{\cal X})^{2}-\frac{2A_{5}}{w_{1}^{2}}\,\frac{d}{dt}(a^{3}Q\,\dot{{\cal R}})\left\{ (\partial_{k}{\cal R})(\partial_{k}{\cal X})-\partial^{-2}\partial_{i}\partial_{j}
 [(\partial_{i}{\cal R})(\partial_{j}{\cal X})] \right\} \nonumber \\
\hspace{-0.3cm} &  &{} +{\rm total~derivatives}\,. 
\label{eq:f1expr2}
\end{eqnarray}
In what follows we omit to write the terms corresponding to total derivatives.
We determine the coefficient $q_{1}$ 
such that the term $-2q_1{\cal R}\dot{{\cal R}}\,\frac{d}{dt}(a^{3}Q\dot{{\cal R}})$
in Eq.~(\ref{eq:f1expr2}) is merged with 
the term $-2aL_1 Q {\cal R}\dot{{\cal R}}\partial^{2}{\cal R}$
in Eq.~(\ref{eq:f2expr2}) to give rise to a combination proportional to the linear
equation of motion for ${\cal R}$ [i.e.\ Eq.~(\ref{linearR2})].
This can be achieved by demanding
\begin{equation}
q_1=-L_1/c_s^2\,.
\label{q1con}
\end{equation}
Since, in general, we have that
\begin{eqnarray}
\hspace{-0.5cm}
c(t)\,\{(\partial_{i}{\cal X})\,(\partial_{i}{\cal R})\,(\partial^{2}{\cal R})-{\cal R}\,\partial_{i}\partial_{j}[(\partial_{i}{\cal R})\,(\partial_{j}{\cal X})]\} 
& = & c(t) \{(\partial_{i}{\cal X})\,(\partial_{i}{\cal R})\,(\partial^{2}{\cal R})-(\partial^{-2}\partial^{2}{\cal R})
\,\partial_{i}\partial_{j}[(\partial_{i}{\cal R})\,(\partial_{j}{\cal X})]\} \nonumber \\
\hspace{-0.5cm}
& = & c(t) \partial^{2}{\cal R}\,\{ (\partial_{i}{\cal R})(\partial_{i}{\cal X})-\partial^{-2}\partial_{i}\partial_{j}[(\partial_{i}{\cal R})\,(\partial_{j}{\cal X})]\}\,,
\end{eqnarray}
the last term in Eq.~(\ref{eq:f1expr2}) can be merged with
the term including $q_2$ in Eq.~(\ref{eq:f2expr2}) as follows:
\begin{eqnarray}
& &\biggl[-\frac{2A_{5}}{w_{1}^{2}}\,\frac{d}{dt}(a^{3}Q\,\dot{{\cal R}})
 +\frac{aq_2}{2}\,\partial^{2}{\cal R}\biggr]\{(\partial_{k}{\cal R})(\partial_{k}{\cal X})-\partial^{-2}\partial_{i}\partial_{j}[(\partial_{i}{\cal R})(\partial_{j}{\cal X})]\}
\nonumber  \\
&&=\frac{A_5}{w_1^2} 
\{(\partial_{k}{\cal R})(\partial_{k}{\cal X})-\partial^{-2}\partial_{i}\partial_{j}[(\partial_{i}{\cal R})(\partial_{j}{\cal X})]\}
\frac{\delta {\cal L}_2}{\delta {\cal R}} \bigg|_{1} \nonumber \\
& &~~~+a\left[\frac{q_2}{2}-\frac{2c_{s}^{2}A_{5}Q}{w_{1}^{2}}\right](\partial^{2}{\cal R})\{(\partial_{k}{\cal R})(\partial_{k}{\cal X})-\partial^{-2}\partial_{i}\partial_{j}[(\partial_{i}{\cal R})(\partial_{j}{\cal X})]\},
\end{eqnarray}
where $\delta {\cal L}_2/\delta {\cal R}|_{1}$ is defined in Eq.~(\ref{linearR}).
Along the same lines one can show that
\begin{eqnarray}
c(t) \{(\partial{\cal R})^{2}\,(\partial^{2}{\cal R})-{\cal R}\,\partial_{i}\partial_{j}[(\partial_{i}{\cal R})\,(\partial_{j}{\cal R})]\} & = & c(t) \{(\partial{\cal R})^{2}\,(\partial^{2}{\cal R})-(\partial^{-2}\partial^{2}{\cal R}) 
\,\partial_{i}\partial_{j}[(\partial_{i}{\cal R})\,(\partial_{j}{\cal R})] \nonumber \\
 & = & c(t) \partial^{2}{\cal R}\,\{(\partial{\cal R})^{2}-\partial^{-2}\partial_{i}\partial_{j}[(\partial_{i}{\cal R})\,(\partial_{j}{\cal R})]\}\,.
\end{eqnarray}
We use this result to merge two different contributions,
namely the term $f_3/a$ in Eq.~(\ref{eq:f3expr2}) and 
the first term on the r.h.s.\ of Eq.~(\ref{eq:f2expr2}).
We have 
\begin{eqnarray}
& &\frac{1}{a^{2}}\biggl[-\frac{1}{2w_{1}}\,(A_{7}-2A_{5}L_{1})\,
\frac{d}{dt}(a^{3}Q\,\dot{{\cal R}})+aq_3 \partial^{2}{\cal R}\biggr]\{(\partial{\cal R})^{2}-\partial^{-2}\partial_{i}\partial_{j}[(\partial_{i}{\cal R})
(\partial_{j}{\cal R})]\} \nonumber \\
 &&=\frac{1}{4w_{1}a^{2}}\,(A_{7}-2A_{5}L_{1})
 \{(\partial{\cal R})^{2}-\partial^{-2}\partial_{i}\partial_{j}[(\partial_{i}{\cal R})(\partial_{j}{\cal R})]\} \frac{\delta {\cal L}_2}{\delta {\cal R}} \bigg|_{1}
 \nonumber \\
 &  &~~~~{}+\frac{1}{a}\left[ q_3-\frac{Qc_{s}^{2}}{2w_{1}}\,(A_{7}-2A_{5}L_{1})\right](\partial^{2}{\cal R})\,\{(\partial{\cal R})^{2}-\partial^{-2}\partial_{i}\partial_{j}[(\partial_{i}{\cal R})(\partial_{j}{\cal R})]\}\,.
\end{eqnarray}

Finally, merging all the contributions together, the third-order action  
(\ref{S3first}) can be written as 
\begin{equation}
S_{3}=\int dt\,{\cal L}_{3}\,,
\end{equation}
where 
\begin{eqnarray}
{\cal L}_{3} &=& \int d^{3}x
\biggl\{ a^3 {\cal C}_{1}\Mpl^2 {\cal R}\dot{{\cal R}}^{2}
+a\,{\cal C}_{2} \Mpl^2 {\cal R}(\partial{\cal R})^{2}
+ a^3{\cal C}_{3}\Mpl \dot{{\cal R}}^{3}
+a^3 {\cal C}_{4} \dot{{\cal R}}(\partial_i{\cal R})(\partial_i{\cal X})+
a^3 ({\cal C}_5/\Mpl^2) \partial^2 {\cal R} (\partial {\cal X})^2
\nonumber \\
& &{} +a{\cal C}_6 \dot{\cal R}^2 \partial^2 {\cal R} 
+({\cal C}_7/a) \left[ \partial^2 {\cal R} (\partial {\cal R})^2
-{\cal R} \partial_i  \partial_j (\partial_i {\cal R}) (\partial_j {\cal R}) \right]
+a({\cal C}_8/\Mpl) \left[ \partial^2 {\cal R} \partial_i {\cal R} \partial_i {\cal X}
-{\cal R}\partial_i  \partial_j (\partial_i {\cal R}) (\partial_j {\cal X}) \right] 
\nonumber \\
& &{}+{\cal F}_1 \frac{\delta {\cal L}_2}{\delta{\cal R}}\biggr|_{1} \biggr\}\,.
\label{L3}
\end{eqnarray}
The dimensionless coefficients ${\cal C}_i$ ($i=1, \cdots, 8$) are given by 
\begin{eqnarray}
\hspace{-0.5cm}{\cal C}_{1} &=& \frac{1}{\Mpl^2} 
\left[ A_{4}+q_{1}(\dot{Q}+3HQ)-Q\dot{q}_{1} \right]
=\frac{Q}{M_{\rm pl}^2} 
\left[ 3-\frac{L_1 H}{c_s^2} \left( 3+\frac{\dot{Q}}{H Q} \right)
+\frac{d}{dt} \left( \frac{L_1}{c_s^2} \right) \right]\,, 
\label{C1f}\\
\hspace{-0.5cm}{\cal C}_{2} &=& \frac{1}{\Mpl^2} 
\left[ A_{8}+\frac{1}{a}\frac{d}{dt}(aL_1 Q)
\right]=\frac{1}{\Mpl^2} \left[\Mpl^2 F-8 \ddot{\xi}+
\frac{1}{a} \frac{d}{dt} \bigl( aL_1 (Q-w_1) \bigr)
\right]\,, \\
\hspace{-0.5cm}{\cal C}_{3} &=& \frac{1}{\Mpl} 
\left( A_1+A_3 \frac{Q}{w_1}-q_1 Q \right)
=\frac{1}{\Mpl}\Biggl\{  L_1 \left[ L_1 (L_1 a_1+a_3)
+a_{15}+(a_{12}+L_1a_{5})\frac{Q}{w_1}
+\frac{Q}{c_s^2} \right]
+8 \dot{\xi}\left(1-\frac{Q}{w_1} \right)
\Biggr\}, \\
\hspace{-0.5cm}{\cal C}_{4} &=& \frac{Q}{w_{1}}\left[
\frac{1}{w_1} (A_6+A_9)
-w_{1}\,\frac{d}{dt}\left(\frac{A_{5}}{w_{1}^{2}}\right)
+\frac{3HA_{5}}{w_{1}}\right]=
-\frac{Q}{2w_1} \left\{
1+2w_1 \left[ \frac{d}{dt} \left( \frac{A_5}{w_1^2} \right)
-3H  \frac{A_5}{w_1^2} \right] \right\}\,, \\
\hspace{-0.5cm}{\cal C}_5 &=& 
\frac{\Mpl^2}{2}\left[\frac{A_{6}}{w_{1}^{2}}-\frac{d}{dt}
\left(\frac{A_{5}}{w_{1}^{2}}\right)+\frac{3HA_{5}}{w_{1}^{2}}\right]
=\frac{\Mpl^2}{2w_1^2} \left[ \frac32 \Mpl^2 
F (1-HL_1)-24 H\dot{\xi} \left(1-\frac32HL_1 \right)
\right]-\frac12 \frac{d}{dt} \left( \frac{A_5}{w_1^2} 
\right) \Mpl^2\,,\\
\hspace{-0.5cm}{\cal C}_6 &=& A_2-A_3 L_1
= L_1 \left[ L_1 (2\Mpl^2 F -L_1a_5
-64 H \dot{\xi})+24 \dot{\xi} \right]\,,\\
\hspace{-0.5cm}{\cal C}_7 &=& 
q_3-\frac{Qc_{s}^{2}}{2w_{1}}\,(A_{7}-2A_{5}L_{1})
=\frac16 L_1^2 \left[ \Mpl^2 F (1-HL_1)
-8H\dot{\xi} (4-3HL_1) \right]-\frac{c_s^2 Q}{2w_1}L_1
\left[ L_1 (\Mpl^2 F -24H\dot{\xi})+16 \dot{\xi} \right] 
\nonumber \\
& &~~~~~~~~~~~~~~~~~~~~~~~~~~~~~~~~~~\,
{}+\frac16 \frac{d}{dt} \left( L_1^2
 \left[ L_1 (\Mpl^2 F-24H\dot{\xi})+24 \dot{\xi}
 \right] \right)\,, \\
\hspace{-0.5cm}{\cal C}_8 &=& \Mpl \left( \frac{q_2}{2}-\frac{2c_s^2 A_5 Q}{w_1^2} \right)=\biggl\{ \frac{L_1}{w_1} (\Mpl^2 F
 -24H\dot{\xi})(HL_1-1)+\frac{c_s^2 Q}{w_1^2}
\left[ L_1 (\Mpl^2 F-24 H\dot{\xi})+8 \dot{\xi} \right] \nonumber \\
&&~~~~~~~~~~~~~~~~~~~~~~~~~~~~~~~~~~
{}-\frac{d}{dt} \biggl[ \frac{L_1[L_1 (\Mpl^2 F-24H \dot{\xi})
+16 \dot{\xi}]}{2w_1} \biggr] \biggr\} \Mpl\,,
\label{C8f}
\end{eqnarray}
with $A_5=-(L_1/2) (M_{\rm pl}^2 F -24H\dot{\xi})-4\dot{\xi}$.
Note that we have used $A_4=3Q$ to simplify ${\cal C}_1$.
The coefficient in front of 
$\delta {\cal L}_2/\delta {\cal R}|_1$ in Eq.~(\ref{L3}) is 
\begin{equation}
{\cal F}_1=
\frac{A_{5}}{w_{1}^{2}}\,\{(\partial_{k}{\cal R})(\partial_{k}{\cal X})
-\partial^{-2}\partial_{i}\partial_{j}[(\partial_{i}{\cal R})(\partial_{j}{\cal X})]\}
+q_{1}{\cal R}\dot{{\cal R}}+\frac{A_{7}-2A_{5}L_{1}}{4w_{1}a^{2}}
\{(\partial{\cal R})^{2}
-\partial^{-2}\partial_{i}\partial_{j}[(\partial_{i}{\cal R})(\partial_{j}{\cal R})]\}\,.
\end{equation}
Each contribution to ${\cal F}_1$ includes 
spatial and time derivatives of
${\cal R}$, which vanish in the large-scale limit ($k \to 0$).
Moreover the term $\delta {\cal L}_2/\delta {\cal R}|_1$ survives
only at second order in ${\cal R}$.
When we evaluate the three-point correlation function of ${\cal R}$ below,
we neglect the contribution of the last term in Eq.~(\ref{L3}) 
relative to those coming from other terms.
In Refs.~\cite{Maldacena,Seery,Chen} the term proportional to ${\cal R}^2$
is present in the definition of ${\cal F}_1$. After a suitable field redefinition
this gives rise to a correction of the order of the slow-roll parameter 
$\eta=\dot{\epsilon}/(H \epsilon)$ in standard inflation \cite{Koyama}.
We have absorbed such a contribution to other terms in Eq.~(\ref{L3})
[such as ${\cal C}_1$]
and hence the field redefinition is not required in our method.

The Hamiltonian in the interacting picture is given by 
${\cal H}_{{\rm int}}=-{\cal L}_{3}$ \cite{Maldacena,Seery,Chen}.
The vacuum expectation value of ${\cal R}$ for the three-point operator
at the conformal time $\tau=\tau_f$ can be expressed as 
\begin{equation}
\langle{\cal R}({\bm{k}}_{1}){\cal R}({\bm{k}}_{2}){\cal R}({\bm{k}}_{3})\rangle
=-i\int_{\tau_i}^{\tau_f}d\tau\, a\,\langle0|\,[{\cal R}(\tau_f,{\bm{k}}_{1})
{\cal R}(\tau_f,{\bm{k}}_{2}){\cal R}(\tau_f,{\bm{k}}_{3}),
{\cal H}_{{\rm int}}(\tau)]\,|0\rangle\,,
\label{Rvacuum}
\end{equation}
where $\tau_i$ is the initial time when the perturbations are 
deep inside the Hubble radius.
Since $\tau \simeq -1/(aH)$ during inflation it is a good approximation 
to take $\tau_i \to -\infty$ and $\tau_f \to 0$, where 
the latter corresponds to some time after 
the Hubble radius crossing.

In order to evaluate the vacuum expectation value (\ref{Rvacuum})
we use the curvature perturbation (\ref{RFourier}) in Fourier space
with the mode function $u(\tau, k)$ given in Eq.~(\ref{usol}).
Each term in the third-order Lagrangian (\ref{L3}) includes the 
phase factor of the form $\int d^3 x\,
e^{i ({\bm k}_4+{\bm k}_5+{\bm k}_6)\cdot {\bm x}}$, 
which gives rise to the delta function 
$(2\pi)^3\,\delta^{(3)} ({\bm k}_4+{\bm k}_5+{\bm k}_6)$.
Among the combination of the annihilation and creation 
operators, the non-vanishing components are
\begin{eqnarray}
& &\langle 0| a({\bm k}_1) a({\bm k}_2) a({\bm k}_3)
a^\dagger (-{\bm k}_4) a^\dagger (-{\bm k}_5) a^\dagger(-{\bm k}_6)|
0 \rangle
=\langle 0| a({\bm k}_4) a({\bm k}_{5}) a({\bm k}_6)
a^\dagger (-{\bm k}_1) a^\dagger (-{\bm k}_2) a^\dagger(-{\bm k}_3)|
0 \rangle \nonumber \\
& &=(2\pi)^9 \biggl\{ \delta^{(3)} ({\bm k}_4+{\bm k}_1) 
\left[\delta^{(3)} ({\bm k}_5+{\bm k}_2)\,\delta^{(3)}({\bm k}_6+{\bm k}_3)+
\delta^{(3)} ({\bm k}_5+{\bm k}_3)\,\delta^{(3)}({\bm k}_6+{\bm k}_2) \right] 
\nonumber \\
& &~~~~~~~~~+\delta^{(3)} ({\bm k}_4+{\bm k}_2) 
\left[\delta^{(3)} ({\bm k}_5+{\bm k}_1)\,\delta^{(3)}({\bm k}_6+{\bm k}_3)+
\delta^{(3)} ({\bm k}_5+{\bm k}_3)\,\delta^{(3)}({\bm k}_6+{\bm k}_1) \right] 
\nonumber \\
& &~~~~~~~~~+\delta^{(3)} ({\bm k}_4+{\bm k}_3) 
\left[\delta^{(3)} ({\bm k}_5+{\bm k}_1)\,\delta^{(3)}({\bm k}_6+{\bm k}_2)+
\delta^{(3)} ({\bm k}_5+{\bm k}_2)\,\delta^{(3)}({\bm k}_6+{\bm k}_1)\right]\biggr\}\,.
\end{eqnarray}

When we carry out the integral in terms of $\tau$ we assume that 
all the terms ${\cal C}_i$ $(i=1, \cdots, 8)$ slowly vary in time 
relative to the scale factor $a$, so that it is a good approximation 
to treat them as constants for the integration.
If this condition is not satisfied, one needs to numerically
solve the integrals without approximations.
In the following we present each contribution of the three-point correlation 
function coming from the integrals given in Eq.~(\ref{L3}). 

\begin{itemize}
\item (1) $H_{{\rm int}}^{(1)}=-\int d^{3}x\,a^{3}{\cal C}_{1}\Mpl^2{\cal R}\dot{{\cal R}}^{2}$
\begin{equation}
\langle{\cal R}({\bm{k}}_{1}){\cal R}({\bm{k}}_{2}){\cal R}({\bm{k}}_{3})\rangle^{(1)}=
(2\pi)^{3}\delta^{(3)}({\bm{k}}_{1}+{\bm{k}}_{2}+{\bm{k}}_{3})
\frac{{\cal C}_{1}\Mpl^2 H^{4}}{16Q^3c_{s}^6}\frac{1}{(k_{1}k_{2}k_{3})^{3}}\left(\frac{k_{2}^{2}k_{3}^{2}}{K}+\frac{k_{1}k_{2}^{2}k_{3}^{2}}{K^{2}}+{\rm sym}\right)\,,
\label{H1}
\end{equation}
where $K=k_{1}+k_{2}+k_{3}$. 
The symbol ``sym'' means the symmetric terms with respect to $k_1, k_2, k_3$.
\item (2) $H_{{\rm int}}^{(2)}=-\int d^{3}x\,a{\cal C}_{2}\Mpl^2 {\cal R}(\partial{\cal R})^{2}$
\begin{eqnarray}
\langle{\cal R}({\bm{k}}_{1}){\cal R}({\bm{k}}_{2}){\cal R}({\bm{k}}_{3})\rangle^{(2)} 
& = & (2\pi)^{3}\delta^{(3)}({\bm{k}}_{1}+{\bm{k}}_{2}+{\bm{k}}_{3})\frac{{\cal C}_{2}\Mpl^2H^{4}}
{16Q^{3}c_{s}^{8}}\frac{1}{(k_{1}k_{2}k_{3})^{3}}\nonumber \\
 &  & \times\left[({\bm{k}}_{1}\cdot{\bm{k}}_{2}+{\bm{k}}_{2}\cdot{\bm{k}}_{3}+{\bm{k}}_{3}\cdot{\bm{k}}_{1})\left(-K+\frac{k_{1}k_{2}+k_{2}k_{3}+k_{3}k_{1}}{K}+\frac{k_{1}k_{2}k_{3}}{K^{2}}\right)\right]\,.
\end{eqnarray}
\item (3) $H_{{\rm int}}^{(3)}=-\int d^{3}x\,a^{3}{\cal C}_{3}\Mpl \dot{{\cal R}}^{3}$
\begin{equation}
\langle{\cal R}({\bm{k}}_{1}){\cal R}({\bm{k}}_{2}){\cal R}({\bm{k}}_{3})\rangle^{(3)} 
=(2\pi)^{3}\delta^{(3)}({\bm{k}}_{1}+{\bm{k}}_{2}+{\bm{k}}_{3})
\frac{3{\cal C}_{3}\Mpl H^{5}}{8Q^3c_{s}^6}\frac{1}{k_{1}k_{2}k_{3}}\frac{1}{K^{3}}\,.
\end{equation}
\item (4) $H_{{\rm int}}^{(4)}=-\int d^{3}x\,a^3 {\cal C}_{4}\dot{{\cal R}}
(\partial_i{\cal R})(\partial_i {\cal X})$
\begin{equation}
\langle{\cal R}({\bm{k}}_{1}){\cal R}({\bm{k}}_{2}){\cal R}({\bm{k}}_{3})\rangle^{(4)}=(2\pi)^{3}\delta^{(3)}({\bm{k}}_{1}+{\bm{k}}_{2}+{\bm{k}}_{3})\frac{{\cal C}_{4}H^{4}}{32Q^{2}c_{s}^{6}}\frac{1}{(k_{1}k_{2}k_{3})^{3}}\left[\frac{({\bm{k}}_{1}\cdot{\bm{k}}_{2})k_{3}^{2}}{K}\left(2+\frac{k_{1}+k_{2}}{K}\right)+{\rm sym}\right]\,.
\end{equation}
\item (5) $H_{{\rm int}}^{(5)}=-\int d^{3}x\, a^3 
({\cal C}_5/\Mpl^2) \partial^2 {\cal R} (\partial {\cal X})^2$
\begin{equation}
\langle{\cal R}({\bm{k}}_{1}){\cal R}({\bm{k}}_{2}){\cal R}({\bm{k}}_{3})\rangle^{(5)}=(2\pi)^{3}\delta^{(3)}({\bm{k}}_{1}+{\bm{k}}_{2}+{\bm{k}}_{3})\frac{{\cal C}_{5}H^{4}}{16Q \Mpl^2 c_{s}^{6}}
\frac{1}{(k_{1}k_{2}k_{3})^{3}}\left[
\frac{k_1^2 ({\bm{k}}_{2} \cdot {\bm{k}}_{3})}{K}
\left(1+\frac{k_1}{K} \right) +{\rm sym} \right]\,.
\end{equation}
\item (6) $H_{{\rm int}}^{(6)}=-\int d^{3}x\,a {\cal C}_6 \dot{\cal R}^2 \partial^2 {\cal R}$
\begin{equation}
\langle{\cal R}({\bm{k}}_{1}){\cal R}({\bm{k}}_{2}){\cal R}({\bm{k}}_{3})\rangle^{(6)}
=(2\pi)^{3}\delta^{(3)}({\bm{k}}_{1}+{\bm{k}}_{2}+{\bm{k}}_{3})
\frac{3{\cal C}_{6}H^{6}}{4Q^{3}c_{s}^{8} }
\frac{1}{k_{1}k_{2}k_{3}}\frac{1}{K^3}\,.
\end{equation}
\item (7) $H_{{\rm int}}^{(7)}=-\int d^{3}x\,
({\cal C}_7/a) \left[ \partial^2 {\cal R} (\partial {\cal R})^2
-{\cal R} \partial_i  \partial_j (\partial_i {\cal R}) (\partial_j {\cal R}) \right]$
\begin{eqnarray}
\langle{\cal R}({\bm{k}}_{1}){\cal R}({\bm{k}}_{2}){\cal R}({\bm{k}}_{3})\rangle^{(7)}
&=& (2\pi)^{3}\delta^{(3)}({\bm{k}}_{1}+{\bm{k}}_{2}+{\bm{k}}_{3})
\frac{{\cal C}_{7}H^{6}}{8Q^{3}c_{s}^{10}}
\frac{1}{(k_{1}k_{2}k_{3})^3}\frac{1}{K}
\left[ 1+\frac{k_1k_2+k_2k_3+k_3k_1}{K^2}+
\frac{3k_1k_2k_3}{K^3} \right] \nonumber \\
&& \times \left[ k_1^2 ({\bm k}_2 \cdot {\bm k}_3)
-({\bm k}_1 \cdot {\bm k}_2)({\bm k}_1 \cdot {\bm k}_3)+{\rm sym} \right]\,.
\end{eqnarray}
\item (8) $H_{{\rm int}}^{(8)}=-\int d^{3}x\,a({\cal C}_8/\Mpl) 
\left[ \partial^2 {\cal R} 
\partial_i {\cal R} \partial_i {\cal X}-{\cal R}\partial_i  \partial_j 
(\partial_i {\cal R}) (\partial_j {\cal X}) \right]$ 
\begin{eqnarray}
\hspace{-0.8cm}
\langle{\cal R}({\bm{k}}_{1}){\cal R}({\bm{k}}_{2}){\cal R}({\bm{k}}_{3})\rangle^{(8)}
&=& (2\pi)^{3}\delta^{(3)}({\bm{k}}_{1}+{\bm{k}}_{2}+{\bm{k}}_{3})
\frac{{\cal C}_{8}H^{5}}{32Q^2\Mpl c_{s}^{8}}
\frac{1}{(k_{1}k_{2}k_{3})^3}\frac{1}{K} \nonumber \\
\hspace{-0.8cm}
&\times& \left\{ \left( 2+\frac{2k_1+k_2+k_3}{K}+
\frac{2k_1 (k_2+k_3)}{K^2} \right) 
\left[ k_1^2 ({\bm k}_2 \cdot {\bm k}_3)
-({\bm k}_1 \cdot {\bm k}_2)({\bm k}_1 \cdot {\bm k}_3) \right]
+{\rm sym} \right\}.
\label{H8}
\end{eqnarray}
\end{itemize}

We write the three-point correlation function of the curvature 
perturbation, as
\begin{equation}
\langle {\cal R} ({\bm k}_1) {\cal R} ({\bm k}_2) {\cal R} ({\bm k}_3)
\rangle =(2\pi)^3  \delta^{(3)} ({\bm k}_1+{\bm k}_2+{\bm k}_3)
B_{\cal R} (k_1, k_2, k_3)\,,
\end{equation}
where 
\begin{equation}
B_{\cal R}(k_1, k_2, k_3)=\frac{(2\pi)^4\,({\cal P}_{\cal R})^2}
{\prod_{i=1}^3 k_i^3}{\cal A}_{\cal R}(k_1, k_2, k_3)\,.
\end{equation}
Recall that the power spectrum ${\cal P}_{\cal R}$ is given by Eq.~(\ref{scalarpower}).
Collecting all the terms in Eqs.~(\ref{H1})-(\ref{H8}), it follows that 
\begin{eqnarray}
\label{AR}
\hspace{-1cm}{\cal A}_{\cal R} &=&
\frac{\Mpl^2}{Q} \Biggl\{ \frac14
\left( \frac{2}{K} \sum_{i>j}k_i^2 k_j^2
-\frac{1}{K^2} \sum_{i \neq j}k_i^2 k_j^3 \right) {\cal C}_1
+\frac{1}{4c_s^2} \left(\frac12 \sum_i k_i^3
+\frac{2}{K} \sum_{i>j} k_i^2 k_j^2-\frac{1}{K^2}
\sum_{i \neq j} k_i^2 k_j^3 \right) {\cal C}_2 \nonumber \\
& &~~~~+\frac32 \frac{H}{\Mpl} \frac{(k_1k_2 k_3)^2}{K^3} 
{\cal C}_3+\frac18 \frac{Q}{\Mpl^2}
\left( \sum_i k_i^3-\frac12 \sum_{i \neq j} k_i k_j^2
-\frac{2}{K^2} \sum_{i \neq j} k_i^2 k_j^3 \right) {\cal C}_4
\nonumber \\
& &~~~~+\frac14 \left( \frac{Q}{\Mpl^2} \right)^2\,\frac1{K^2}
\left[ 
\sum_i k_i^5+\frac12\sum_{i\neq j}k_i k_j^4-\frac32\sum_{i\neq j} k_i^2k_j^3-k_1k_2k_3\sum_{i>j} k_i k_j
\right]{\cal C}_5
+\frac{3}{c_s^2} \left(\frac{H}{\Mpl} \right)^2 
 \frac{(k_1k_2 k_3)^2}{K^3} {\cal C}_6 \nonumber \\
 & &~~~~+\frac{1}{2c_s^4} \left(\frac{H}{\Mpl} \right)^2 
\frac{1}{K} \left( 1+\frac{1}{K^2}\,\sum_{i>j}k_ik_j+
\frac{3k_1k_2k_3}{K^3} \right) 
\left[
\frac34\,\sum_i k_i^4-\frac32\sum_{i>j}k_i^2k_j^2
\right]
\,{\cal C}_7 \nonumber \\
 & &~~~~+\frac{1}{8c_s^2}\frac{H}{\Mpl} \frac{Q}{\Mpl^2}
\frac{1}{K^2} 
\left[
\frac32\,k_1k_2k_3\sum_i k_i^2-\frac52\,k_1k_2k_3K^2-6\sum_{i\neq j}k_i^2k_j^3-\sum_i k_i^5+\frac72\,K\sum_i k_i^4
\right]{\cal C}_8 
\Biggr\}\,,
\end{eqnarray}
which has an explicit dependence on the wave numbers.
The bispectrum (\ref{AR}) is the central result of our work.

In general the non-linear parameter $f_{\rm NL}$ associated 
with non-Gaussianities of the curvature perturbation 
is defined as \cite{WMAP1and5,WMAP7,Taka}
\begin{equation}
f_{\rm NL}=\frac{10}{3} \frac{{\cal A}_{\cal R}}
{\sum_{i=1}^3 k_i^3}\,,
\end{equation}
which matches with the notation of the WMAP group.
For the equilateral configuration with $k_1=k_2=k_3$ one has 
\begin{eqnarray}
f_{\rm NL}^{\rm equil} &=&
\frac{40}{9} \frac{\Mpl^2}{Q}
\biggl[ \frac{1}{12} {\cal C}_1+\frac{17}{96c_s^2}{\cal C}_2
+\frac{1}{72} \frac{H}{\Mpl}{\cal C}_3-\frac{1}{24} 
\frac{Q}{\Mpl^2}{\cal C}_4-\frac{1}{24} 
\left( \frac{Q}{\Mpl^2} \right)^2 {\cal C}_5
+\frac{1}{36c_s^2} \left( \frac{H}{\Mpl} \right)^2 {\cal C}_6
\nonumber \\
& &~~~~~~~~~~
-\frac{13}{96 c_s^4} \left( \frac{H}{\Mpl} \right)^2 
{\cal C}_7-\frac{17}{192c_s^2} \frac{H}{\Mpl}
\frac{Q}{\Mpl^2}{\cal C}_8 \biggr]\,,
\label{fnl}
\end{eqnarray}
which is independent of the wave numbers.
Recall that the coefficients ${\cal C}_i$ ($i=1,\cdots,8$)
can be evaluated by Eqs.~(\ref{C1f})-(\ref{C8f}).

\section{Expansion in terms of slow variation parameters}
\label{expansion} 

Let us derive a simpler form of $f_{\rm NL}^{\rm equil}$
by using the approximation that the slow-variation 
terms defined in Eq.~(\ref{slowvariation}) are much 
smaller than 1.
We also use the expansion of $L_1$ and $\epsilon_s$
given in Eqs.~(\ref{L1expansion}) and (\ref{epsilons}), respectively.
First we write the variables $Q$ and $c_s^2$ 
in the forms
\begin{equation}
Q=\frac{4\Mpl^4 \Sigma}{w_2^2}\,,\qquad
c_s^2=\frac{2 w_1^{2} w_2H-w_2^2 w_4
+4 w_1 \dot{w}_1w_2
-2w_1^{2}\dot{w}_2}{4\Mpl^4 \Sigma}\,,
\label{Qf}
\end{equation}
where 
\begin{equation}
\Sigma \equiv \frac{w_1 (4w_1w_3+9w_2^2)}
{12 \Mpl^4}\,.
\label{Sigmadef}
\end{equation}
We also define the following quantities
\begin{equation}
\lambda \equiv F^2 \left[ X^2 P_{,XX}+2X^3 P_{,XXX}/3
+\dot{\phi}H (XG_{,X}+5X^2 G_{,XX}+2X^3 G_{,XXX})
-2(2X^2 G_{,\phi X}+X^3 G_{,\phi XX})/3 \right]\,,
\end{equation}
and 
\begin{equation}
\lambda_{G}\equiv X\,G_{,XX}/G_{,X}\,.
\end{equation}
It is convenient to introduce $\Sigma$ and $\lambda$
in order to obtain a compact expression of 
$f_{\rm NL}^{\rm equil}$.
Note that these parameters have been already introduced
in the context of k-inflation \cite{Seery,Chen}. 

We first replace $w_{3}$ with respect to $\Sigma$, which in turn can be
written in terms of $\epsilon_s$. 
Using Eq.~(\ref{epsilonsdef}), the parameter $\epsilon_s$ can be expressed 
in terms of other slow-variation parameters.
Then we replace $F$ as $F=Qc_{s}^{2}/(\epsilon_s M_{{\rm pl}}^{2})$. Finally we expand the coefficients ${\cal C}_i$ ($i=1,\cdots,8$)
in terms of the slow-variation parameters, 
by keeping $c_{s}^{2}$ and $\lambda_G$ as 
unknown parameters. A similar method is applied to all the other terms,
except for ${\cal C}_{3}$. In this case the terms $P_{,X}$
and $G_{,XXX}$ are expressed in terms of $\Sigma$ and $\lambda$, 
respectively. We replace any left $G_{,XX}$ with $\lambda_{G}$. We also multiply and divide $\lambda$ by $\Sigma$, and we consider the ratio $\lambda/\Sigma$ as a free parameter of the theories (as we do for $c_s^2$ and $\lambda_G$). Then, as done for the other terms, we replace $\Sigma$ in terms of $Q$, and any left $F$ in terms of $Q$ and $\epsilon_s$. Finally we use the relation between $\epsilon$ and
other slow-variation parameters.
By following this procedure, the final expression of each 
$f_{{\rm NL}}^{{\rm equil}\,(i)}$ ($i=1, \cdots, 8$) only depends on 
$\epsilon_s$, $\eta_s$, $s$, $\delta_F$, $\delta_{\xi}$, $\delta_{GX}$, 
and the free parameters $c_s^2$, $\lambda_G$, and $\lambda/\Sigma$.
This form allows an expansion with respect to the slow-variation parameters.
Then the leading contributions to each $f_{{\rm NL}}^{{\rm equil}\,(i)}$
coming from the coefficients ${\cal C}_i$ are given by 
\begin{eqnarray}
f_{{\rm NL}}^{{\rm equil}\,(1)} & = & \frac{10}{9}\left(1-\frac{1}{c_{s}^{2}}\right)-\frac{10}{27c_{s}^{2}}\left(8\,\delta_{{\xi}}+\eta_{s}+4\,\delta_{{GX}}-\epsilon_s\right),\label{eq:fnl1approx}\\
f_{{\rm NL}}^{{\rm equil}\,(2)} & = & \frac{85}{108}\left(\frac{1}{c_{s}^{2}}-1\right)+\frac{85}{108c_{s}^{2}}\left(\epsilon_{s}+8\,\delta_{{\xi}}-2\, s+\eta_{s}\right),\\
f_{{\rm NL}}^{{\rm equil}\,(3)} & = & \frac{5}{81}\left(\frac{1}{c_{s}^{2}}-1\right)-\frac{10}{81}\,\frac{\lambda}{\Sigma}+\frac{5}{162c_{s}^{2}}\left(8\,\delta_{{\xi}}+2\,\delta_{{GX}}-\delta_{{F}}\right)-\frac{5}{162}\left(6\,\delta_{{GX}}-\delta_{{F}}+4\,\lambda_{{G}}\,\delta_{{GX}}+8\,\delta_{{\xi}}\right) \nonumber \\
&&{}+\frac5{27}\,\frac{\delta_{{GX}}}{\epsilon_s}\,c_s^2 \left( 1+\lambda_{{G}} \right) \left( \delta_{{F}}-8\,\delta_{{\xi}}-2\,\delta_{{GX}} \right)\,, \\
f_{{\rm NL}}^{{\rm equil}\,(4)} & = & \frac{10}{27}\,\frac{\epsilon_s}{c_{s}^{2}}\,,\\
f_{{\rm NL}}^{{\rm equil}\,(5)} & = & -\frac{5}{108c_{s}^{2}}\left(\epsilon_{s}-4\,\delta_{{GX}}+16\,\delta_{{\xi}}\right)\epsilon_{s},\\
f_{{\rm NL}}^{{\rm equil}\,(6)} & = & \frac{20}{81}\,\frac{(1+\lambda_{G})\,
\delta_{GX}}{\epsilon_s}\,,\\
f_{{\rm NL}}^{{\rm equil}\,(7)} & = & \frac{65}{162c_{s}^{2}}\,\frac{\delta_{GX}}{\epsilon_s}\,,\\
f_{{\rm NL}}^{{\rm equil}\,(8)} & = & -\frac{85}{108}\,\frac{\delta_{GX}}{c_{s}^{2}}\,.
\label{eq:fnl8approx}
\end{eqnarray}

Let us now add up all the contributions together. 
In doing so we take the largest contributions, that is, we discard
the corrections of the order of slow-variation parameters relative 
to other existent terms. For example, if a term 
$\delta_{GX}/(\epsilon_s  c_{s}^{2})$
already exists, we ignore the terms 
like $\delta_{GX}/c_{s}^{2}$.
Then we have that
\begin{eqnarray}
f_{{\rm NL}}^{{\rm equil}} &\simeq& 
\frac{85}{324}\left(1-\frac{1}{c_{s}^{2}}\right)-\frac{10}{81}\,\frac{\lambda}{\Sigma}+\frac{55}{36}\,\frac{\epsilon_s}{c_{s}^{2}}
+\frac{5}{12}\,\frac{\eta_{s}}{c_{s}^{2}}
-\frac{85}{54}\,\frac{s}{c_{s}^{2}} \nonumber \\
& &+\frac{5}{162}\,\delta_{F}\left(1-\frac{1}{c_{s}^{2}}\right)
-\frac{10}{81}\,\delta_{\xi}\left(2-\frac{29}{c_{s}^{2}}\right)
+\delta_{GX}\left[\frac{20\,(1+\lambda_{G})}{81\epsilon_s}
+\frac{65}{162c_{s}^{2}\epsilon_s}\right]\,,
\label{fnleq}
\end{eqnarray}
where we have tacitly assumed that $c_s^2<{\cal O}(1)$.
The result (\ref{fnleq}) is valid for any quasi de Sitter background.

\section{Concrete models}
\label{cmodels} 

In this section we apply the results derived in previous sections
to concrete models of inflation.
This includes (i) k-inflation, (ii) generalized Galileon model, and 
(iii) models based on BD theories.

\subsection{k-inflation}

Let us first consider k-inflation models with $F=1$, $\xi=0$, 
and $G=0$. Since $\epsilon_s=\epsilon=-\dot{H}/H^2=XP_{,X}/(\Mpl^2 H^2)$ 
and $L_1=1/H$ in those models, it is possible to derive 
an exact expression of $f_{\rm NL}^{\rm equil}$
without employing the approximation 
in terms of slow-variation parameters.
The coefficients (\ref{C1f})-(\ref{C8f}) are given by 
\begin{eqnarray}
& &\C_1=\frac{\epsilon}{c_s^4} (\epsilon-3+3c_s^2-\eta)\,,\qquad
\C_2=\frac{\epsilon}{c_s^2} (1-c_s^2+\epsilon+\eta-2s)\,,\qquad
\C_3=-\frac{1}{H^3 \Mpl} \left[ \left(1-\frac{1}{c_s^2}
\right) \Sigma+2\lambda \right]\,, \nonumber \\
& &\C_4=\frac{\epsilon}{c_s^2} \left(-2+\frac12 \epsilon \right)\,,
\qquad \C_5=\frac14 \epsilon\,,\qquad
\C_6=\C_7=\C_8=0\,,
\label{Ccon}
\end{eqnarray}
where $\eta \equiv \eta_s=\dot{\epsilon}/(H\epsilon)$,
$\Sigma=XP_{,X}+2X^2 P_{,XX}$ and 
$\lambda=X^2 P_{,XX}+2X^3 P_{,XXX}/3$.

The field propagation speed squared is
\begin{equation}
c_s^2=\frac{P_{,X}}{P_{,X}+2XP_{,XX}}
=\frac{XP_{,X}}{\Sigma}=\frac{\Mpl^2 H^2 \epsilon}
{\Sigma}\,.
\end{equation}
Plugging the coefficients (\ref{Ccon}) into Eq.~(\ref{fnl}), 
it follows that 
\begin{equation}
f_{\rm NL}^{\rm equil}
=\frac{85}{324} \left( 1- \frac{1}{c_s^2} \right)
-\frac{10}{81} \frac{\lambda}{\Sigma}
+\frac{55}{36} \frac{\epsilon}{c_s^2}+
\frac{5}{12} \frac{\eta}{c_s^2}-\frac{85}{54} 
\frac{s}{c_s^2}\,,
\label{fnlkinf}
\end{equation}
where we have ignored the second-order term $\epsilon^2/(2c_s^2)$
in the expression of ${\cal C}_4$.
The non-linear parameter (\ref{fnleq}) derived under the slow-variation 
approximation also gives the same result.
Note that this coincides with the result in Refs.~\cite{Seery,Chen}.

In standard inflation driven by a potential energy $V(\phi)$ of 
the field $\phi$, i.e. $P=X-V(\phi)$, one has $c_s^2=1$, 
$\lambda=0$, and $s=0$. Then the equilateral non-linear 
parameter (\ref{fnlkinf}) reduces to 
\begin{equation}
f_{\rm NL}^{\rm equil}
=\frac{55}{36} \epsilon+
\frac{5}{12} \eta\,,
\end{equation}
which means that $|f_{\rm NL}^{\rm equil}| \ll 1$.

\subsection{Generalized Galileon model}

Let us consider the generalized Galileon model with 
$F=1$ and $\xi=0$.
Since the exact formula of 
$f_{\rm NL}^{\rm equil}$ is complicated,
we employ the result (\ref{fnleq}) derived under the slow-variation 
approximation:
\begin{equation}
f_{{\rm NL}}^{{\rm equil}} \simeq
\frac{85}{324}\left(1-\frac{1}{c_{s}^{2}}\right)
-\frac{10}{81}\,\frac{\lambda}{\Sigma}+\frac{55}{36}\,
\frac{\epsilon_s}{c_{s}^{2}}
+\frac{5}{12}\,\frac{\eta_{s}}{c_{s}^{2}}
-\frac{85}{54}\,\frac{s}{c_{s}^{2}}
+\frac{\delta_{GX}}{\epsilon_s}
\left[\frac{20\,(1+\lambda_{G})}{81}+
\frac{65}{162c_{s}^{2}}\right]\,.
\label{fnlGalileon}
\end{equation}
Using the background equation (\ref{dotHeq}), 
the term $\delta_{GX}/\epsilon_s$ can be expressed as
\begin{equation}
\frac{\delta_{GX}}{\epsilon_s} \simeq 
\frac{\delta_{GX}}{\epsilon+\delta_{GX}}
=\frac{H \dot{\phi}\,G_{,X}}
{P_{,X}+(4H\dot{\phi}-\ddot{\phi})G_{,X}-2G_{,\phi}}\,.
\end{equation}
In the limit that $c_s^2 \ll 1$ the dominant contribution 
to $f_{{\rm NL}}^{{\rm equil}}$ is
\begin{equation}
f_{{\rm NL}}^{{\rm equil}} \simeq
-\frac{85}{324} \frac{1}{c_s^2}
-\frac{10}{81}\,\frac{\lambda}{\Sigma}
+\frac{H \dot{\phi}\,G_{,X}}
{P_{,X}+(4H\dot{\phi}-\ddot{\phi})G_{,X}-2G_{,\phi}}
\left[\frac{20\,(1+\lambda_{G})}{81}+
\frac{65}{162c_{s}^{2}}\right]\,.
\label{fnlGalileon2}
\end{equation}
This matches with the result of Ref.~\cite{Mizuno}
in which the authors ignored the terms 
$\ddot{\phi}\,G_{,X}$ and $G_{,\phi}$ relative 
to $H \dot{\phi}\, G_{,X}$. 
The last term in the parenthesis of Eq.~(\ref{fnlGalileon2})
gives rise to an additional contribution to the terms appearing 
in k-inflation.

\subsection{Brans-Dicke theories}

The action in BD theory with a field potential 
$V(\phi)$ is given by 
\begin{equation}
S=\int d^4 x \sqrt{-g} \left[ \frac12 M_{\rm pl} \phi R
+\frac{M_{\rm pl}}{\phi} \omega_{\rm BD} X -V(\phi)
\right]\,,
\end{equation}
where $\omega_{\rm BD}$ is the BD parameter.
Compared to the original paper \cite{BDtheory} we have introduced
the reduced Planck mass $M_{\rm pl}$, so that the field $\phi$
has a dimension of mass.
In this theory the background equation (\ref{dotHeq}) gives 
\begin{equation}
2\epsilon_1=-\epsilon_2+\epsilon_2 \epsilon_3
+\omega_{\rm BD}\, \epsilon_2^2\,,
\end{equation}
where 
\begin{equation}
\epsilon_1 \equiv -\frac{\dot{H}}{H^2}\,,\qquad
\epsilon_2 \equiv \frac{\dot{\phi}}{H \phi}\,,\qquad
\epsilon_3 \equiv \frac{\ddot{\phi}}{H \dot{\phi}}\,.
\end{equation}

Let us derive a full expression of $f_{{\rm NL}}^{{\rm equil}}$
without using the slow-variation approximation.
{}From Eqs.~(\ref{eq:defQ}) and (\ref{eq:defc2s}) it follows that 
\begin{equation}
Q=\frac{(3+2\omega_{\rm BD})\epsilon_2^2 \phi M_{\rm pl}}
{(2+\epsilon_2)^2}\,,\qquad c_s^2=1\,.
\end{equation}
The time-derivatives appearing in the coefficients $C_i$ 
can be expressed in terms of $\epsilon_i$ ($i=1,2,3$) 
by using the relation 
\begin{equation}
\dot{\epsilon}_2=H \epsilon_2 
(\epsilon_1-\epsilon_2+\epsilon_3)\,.
\end{equation}
Then the coefficients $C_i$ ($i=1, \cdots, 8$)
are given by 
\begin{eqnarray}
&&\C_1=-\frac{\phi}{M_{\rm pl}}\,\frac{(3+2\omega_{\rm BD})\epsilon_2^2
\left[(2\omega_{\rm BD}-3)\epsilon_2^2+4(\epsilon_3-3)\epsilon_2+8\epsilon_3\right]}
{(2+\epsilon_2)^4}\,,\\
& &\C_2=\frac{\phi}{M_{\rm pl}} 
\frac{(3+2\omega_{\rm BD})\epsilon_2^2
\left[ 8 \epsilon_3+(4\epsilon_3-12) \epsilon_2
+(3+6\omega_{\rm BD})\epsilon_2^2\right]}{(2+\epsilon_2)^4}\,,\\
& &\C_3=\C_6=\C_7=\C_8=0\,,\\
& &\C_4=\frac{(3+2\omega_{\rm BD})\epsilon_2^2
\left[-16-16\epsilon_2+(2\omega_{\rm BD}-1)\epsilon_2^2 
\right]}{2(2+\epsilon_2)^4}\,,\\
& &\C_5=\frac{M_{\rm pl}}{\phi}
\frac{(3+2\omega_{\rm BD})\epsilon_2^2}
{4(2+\epsilon_2)^2}\,,
\end{eqnarray}
which lead to 
\begin{equation}
f_{\rm NL}^{\rm equil}
= \frac{5}{18} \frac{48 \epsilon_3-72 (1-\epsilon_3)
\epsilon_2-(6-68 \omega_{\rm BD}-36\epsilon_3) \epsilon_2^2
+(48+68\omega_{\rm BD}+6\epsilon_3) \epsilon_2^3
+(12+11\omega_{\rm BD}-2\omega_{\rm BD}^2)
\epsilon_2^4}{(2+\epsilon_2)^4}.
\end{equation}
Since $|\epsilon_2| \ll 1$ and 
$|\epsilon_3| \ll 1$ during inflation, we have 
\begin{equation}
f_{\rm NL}^{\rm equil} \simeq -\frac{5}{4}\epsilon_2
+\frac{5}{6} \epsilon_3\,,
\label{fnlBD}
\end{equation}
which gives $|f_{\rm NL}^{\rm equil}| \ll 1$.
It is worth mentioning that the same result as Eq.~(\ref{fnlBD})
follows by using the approximated expression of
$f_{\rm NL}^{{\rm equil}\,(i)}$ ($i=1, \cdots, 8$)
derived in Sec.\,\ref{expansion}.
The above analysis covers the case of $f(R)$ gravity in which 
the BD parameter is $\omega_{\rm BD}=0$.
Hence, the Starobinsky's inflation model, $f(R)=R+R^2/(6M^2)$, 
leads to $|f_{\rm NL}^{\rm equil}| \ll 1$ as in standard inflation.

\subsection{Potential driven inflation in the presence
of the nonminimal coupling and the Gauss-Bonnet term}

The result (\ref{fnleq}) shows that, in the limit $c_s^2 \ll 1$,  
the terms coming from the nonminimal coupling 
and the Gauss-Bonnet coupling are
suppressed relative to the term proportional to $1/c_s^2$.
The effect of those couplings on $f_{\rm NL}^{\rm equil}$
appears indirectly through the change of $c_s^2$.

Let us consider the following action
\begin{equation}
S=\int d^4 x \sqrt{-g} \left[ \frac{M_{\rm pl}^2}{2} F(\phi) R
+\omega (\phi) X -V(\phi)-\xi(\phi) {\cal G} \right]\,,
\end{equation}
which corresponds to inflation driven by the field 
potential $V(\phi)$.
Expanding the scalar propagation speed (\ref{eq:defc2s}) up to 
second order, we obtain
\begin{equation}
c_s^2 \simeq 1-\frac{2\delta_{\xi} (\delta_F-8 \delta_{\xi})
(3\delta_F-24\delta_{\xi}-4\delta_{PX})}
{\delta_{PX}}\,,
\end{equation}
where $\delta_{PX}=\omega X/(\Mpl^2 H^2 F)$ is another 
slow-variation parameter.
Since $\{|\delta_{\xi}|,  |\delta_F|, |\delta_{PX}| \} \ll 1$, 
the scalar propagation speed is very close to 1, i.e.
$|c_s^2-1|={\cal O}(\epsilon^2)$.
{}From Eq.~(\ref{fnleq}) we have 
\begin{equation}
f_{\rm NL}^{\rm equil} \simeq \frac{55}{36} \epsilon_s
+\frac{5}{12} \eta_s+\frac{10}{3} \delta_{\xi}\,,
\end{equation}
and hence the non-Gaussianity is small.

The non-Gaussianity can be large for the kinetically driven 
inflation in the presence of the terms 
$P(\phi, X)$, $\xi(\phi){\cal G}$, and 
$G(\phi, X) \square \phi$. 
We will study such models in a separate 
publication \cite{DETT}.

\section{Conclusions}
\label{conclusions}

For the general single-field models described by 
the action (\ref{action}) we have calculated the three-point 
correlation function of curvature perturbations ${\cal R}$
generated during inflation.
This covers the inflationary models motivated by low-energy 
effective string theory, scalar-tensor theories, and Galileon-inspired gravity.
The Gauss-Bonnet term $\xi (\phi) {\cal G}$ and the generalized 
Galileon term $G(\phi,X) \square \phi$ give rise to the 
scalar propagation speed $c_s$ different from 1, 
as it happens in k-inflation.
These models can lead to large non-Gaussianities of primordial
perturbations detectable in future observations.

Using the ADM metric in uniform-field gauge ($\delta \phi=0$), we have 
expanded the action (\ref{action}) up to third order in the perturbations.
The second-order action associated with linear curvature perturbations
is given by Eq.~(\ref{eq:az2}), where both $Q$ and $c_s^2$ 
need to be positive to avoid the appearance
of ghosts and Laplacian instabilities respectively.
In the quasi de Sitter background we have solved the equation 
for curvature perturbations and derived the scalar 
power spectrum and its spectral index. 
The tensor spectrum as well as the tensor-to-scalar ratio is 
also evaluated to confront the models with observations.

After the derivation of the third-order perturbed action (\ref{eq:S3}),
we have expressed it in a more convenient form for the calculation 
of non-Gaussianities by making a lot of integrations by parts.
The coefficient ${\cal F}_1$ in front of $\delta {\cal L}_2/\delta {\cal R}|_1$
includes only the derivative terms of ${\cal R}$, which vanish in the large-scale
limit. Moreover the term $\delta {\cal L}_2/\delta {\cal R}|_1$ survives only at 
second order of perturbations.
When the vacuum expectation value of ${\cal R}$ for the three-point operator 
is evaluated, we just need
to compute the contributions coming from the 8 terms in Eq.~(\ref{L3})
by treating the coefficients $C_i$ ($i=1, \cdots, 8$) 
as slowly varying parameters relative to the scale factor.
Finally we have derived the three-point correlation function as in the form (\ref{AR}).
For the equilateral configuration ($k_1=k_2=k_3$) the non-linear parameter
$f_{\rm NL}^{\rm equil}$ reduces to Eq.~(\ref{fnl}), which
does not have any momentum dependence.

Under the approximation in terms of slow-variation parameters
we showed that $f_{\rm NL}^{\rm equil}$ reduces to a simple
form (\ref{fnleq}). Provided that $c_s^2 \ll 1$, one can realize
large non-Gaussianities with $|f_{\rm NL}^{\rm equil}| \gg 1$.
For $c_s^2 \ll 1$ the terms involving $\delta_F$ and $\delta_{\xi}$
are sub-dominant contributions with respect
to the first two terms in Eq.~(\ref{fnleq}), but 
the last term in Eq.~(\ref{fnleq}), coming from 
the Galileon-type self interaction, can be comparable to 
the dominant contributions.
Compared to k-inflation the presence of the terms $\xi (\phi) {\cal G}$
and $G(\phi, X)\square \phi$ leads to different values of $c_s$, 
so that $f_{\rm NL}^{\rm equil}$ is subject to change.

We have applied our results of the equilateral non-linear parameter
to a number of concrete models of inflation.
In k-inflation and in the generalized Galileon model, 
our formula of $f_{\rm NL}^{\rm equil}$ reproduces
the previous results known in literature.
In the inflationary models based on Brans-Dicke theories, 
including $f(R)$ gravity, the non-linear parameter is of the 
order of slow-roll parameters.
This is associated with the fact that the scalar propagation 
speed is unity in those models.

It will be of interest to distinguish between various inflationary models 
observationally by using our formula of $f_{\rm NL}^{\rm equil}$
as well as $n_{\cal R}$ and $r$.
Especially the potential detectability of non-Gaussianities by the 
PLANCK satellite may open up a new opportunity to approach 
the origin of inflation.

\section*{ACKNOWLEDGEMENTS}
\label{acknow}
We thank Kazuya Koyama, Shuntaro Mizuno, and Takahiro Tanaka 
for useful discussions.
The work of A.\,D.\,F.\ and S.\,T.\ was supported by the Grant-in-Aid
for Scientific Research Fund of the JSPS Nos.~10271 and 30318802.
S.\,T.\ also thanks financial support for the Grant-in-Aid for Scientific
Research on Innovative Areas (No.~21111006).

\appendix

\section{Explicit expression of $Q$ and $c_s$}\label{app:actio3}

Using the expressions for $w_1$, $w_2$, $w_3$ and $w_4$, 
it is possible to write down explicitly the form of $Q$ 
defined in Eq.~(\ref{eq:defQ}). Namely we have
\begin{eqnarray}
Q & \equiv & \frac{w_{1}}{w_{2}^{2}}\,[3\,{M_{{\rm Pl}}}^{4}{\dot{F}}^{2}-2\,{\dot{\phi}}^{4}{M_{{\rm Pl}}}^{2}FG_{,\phi X}-4\,{M_{{\rm Pl}}}^{2}FG_{,\phi}{\dot{\phi}}^{2}+2\, P_{,X}{\dot{\phi}}^{2}{M_{{\rm Pl}}}^{2}F-48\,{\dot{\phi}}^{5}{H}^{2}G_{,XX}\dot{\xi}\nonumber \\
 &  &{} +2\,{\dot{\phi}}^{4}P_{,XX}{M_{{\rm Pl}}}^{2}F+12\,{\dot{\phi}}^{3}G_{,X}{M_{{\rm Pl}}}^{2}FH+6\,{\dot{\phi}}^{5}HG_{,XX}{M_{{\rm Pl}}}^{2}F-48\,{M_{{\rm Pl}}}^{2}\dot{F}{H}^{2}\dot{\xi}\nonumber \\
 &  &{} -6\,{M_{{\rm Pl}}}^{2}\dot{F}{\dot{\phi}}^{3}G_{,X}-48\,{H}^{2}\dot{\xi}{\dot{\phi}}^{3}G_{,X}+16\,{\dot{\phi}}^{4}H\dot{\xi}G_{,\phi X}-16\, P_{,X}{\dot{\phi}}^{2}H\dot{\xi}\nonumber \\
 &  &{} -16\, P_{,X}{\dot{\phi}}^{2}H\dot{\xi}+192\,{H}^{4}{\dot{\xi}}^{2}+32\,{\dot{\phi}}^{2}H\dot{\xi}G_{,\phi}+3\,{\dot{\phi}}^{6}{G_{,X}}^{2}-16\,{\dot{\phi}}^{4}P_{,XX}H\dot{\xi}]\,.
\end{eqnarray}
Along the same lines the speed of propagation $c_s^2$ 
given in Eq.~(\ref{eq:defc2s}) can be written as
\begin{equation}
c_{s}^{2}=\frac{L_{3}\,\ddot{\phi}+L_{4}}{L_{2}}\,,
\end{equation}
where 
\begin{eqnarray}
L_{2} & = & (M_{{\rm Pl}}^{2}F-8H\,\xi_{,\phi}\dot{\phi})\,[3\,{M_{{\rm Pl}}}^{4}{F_{,\phi}}^{2}-2\,{M_{{\rm Pl}}}^{2}F\,{\dot{\phi}}^{2}G_{{,\phi X}}-4\,{M_{{\rm Pl}}}^{2}F\, G_{{,\phi}}+2\,{M_{{\rm Pl}}}^{2}F\, P_{{,X}}\nonumber \\
 &  &{} +2\,{M_{{\rm Pl}}}^{2}F\,{\dot{\phi}}^{2}P_{{,XX}}+12\,{M_{{\rm Pl}}}^{2}F\,\dot{\phi}\, H\, G_{{,X}}+6\,{M_{{\rm Pl}}}^{2}F{\dot{\phi}}^{3}HG_{{,XX}}-48\,{M_{{\rm Pl}}}^{2}F_{{,\phi}}{H}^{2}\xi_{{,\phi}}\nonumber \\
 &  & {}-6\,{M_{{\rm Pl}}}^{2}F_{{,\phi}}{\dot{\phi}}^{2}G_{{,X}}-48\,{H}^{2}\xi_{{,\phi}}{\dot{\phi}}^{2}G_{{,X}}+16\,{\dot{\phi}}^{3}H\xi_{{,\phi}}G_{{,\phi X}}-16\, P_{{,X}}\dot{\phi}\, H\,\xi_{{,\phi}}+192\,{H}^{4}{\xi_{{,\phi}}}^{2}\nonumber \\
 &  & {}+32\,\dot{\phi}\, H\,\xi_{{,\phi}}G_{{,\phi}}+3\,{\dot{\phi}}^{4}{G_{{,X}}}^{2}-16\,{\dot{\phi}}^{3}P_{{,XX}}H\,\xi_{{,\phi}}-48\,{\dot{\phi}}^{4}{H}^{2}G_{{,XX}}\xi_{{,\phi}}]\,,\\
L_{3} & = & 1536\,{H}^{4}{\xi_{{,\phi}}}^{3}+640\,{H}^{2}{\xi_{{,\phi}}}^{2}{\dot{\phi}}^{2}G_{{,X}}+128\,{H}^{2}{\xi_{{,\phi}}}^{2}{\dot{\phi}}^{4}G_{{,XX}}+2\,{M_{{\rm Pl}}}^{4}{F}^{2}{\dot{\phi}}^{2}G_{{,XX}}\nonumber \\
 &  & {}-384\,{M_{{\rm Pl}}}^{2}{\xi_{{,\phi}}}^{2}F_{{,\phi}}{H}^{2}-48\,{M_{{\rm Pl}}}^{2}\xi_{{,\phi}}{\dot{\phi}}^{2}F_{{,\phi}}G_{{,X}}+24\,{\dot{\phi}}^{4}{G_{{,X}}}^{2}\xi_{{,\phi}}+4\,{M_{{\rm Pl}}}^{4}{F}^{2}G_{{,X}}\nonumber \\
 &  & {}+24\,{M_{{\rm Pl}}}^{4}{F_{{,\phi}}}^{2}\xi_{{,\phi}}-32\,{M_{{\rm Pl}}}^{2}F\, H\,\xi_{{,\phi}}{\dot{\phi}}^{3}G_{{,XX}}-64\,{M_{{\rm Pl}}}^{2}F\, H\,\xi_{{,\phi}}\dot{\phi}\, G_{{,X}}\,,\\
L_{4} & = & 8\,{G_{{,X}}}^{2}\xi_{{,\phi\phi}}\,\dot{\phi}^{6}-48\,{G_{{,X}}}^{2}H\xi_{{,\phi}}\,\dot{\phi}^{5}+[32\,\xi_{{,\phi}}G_{{,X}}G_{{,\phi}}-{M_{{\rm Pl}}}^{2}F{G_{{,X}}}^{2}+256\,\xi_{{,\phi}}G_{{,X}}{H}^{2}\xi_{{,\phi\phi}}\nonumber \\
 &  & {}-16\,{M_{{\rm Pl}}}^{2}F_{{,\phi}}G_{{,X}}\xi_{{,\phi\phi}}-16\,\xi_{{,\phi}}G_{{,X}}P_{{,X}}+128\,{H}^{2}{\xi_{{,\phi}}}^{2}G_{{,\phi X}}-16\,{M_{{\rm Pl}}}^{2}\xi_{{,\phi}}G_{{,X}}F_{{,\phi\phi}}]\,\dot{\phi}^{4}\nonumber \\
 &  & {}-32\,\xi_{,\phi}H\left(8\,\xi_{,\phi}G_{,X}{H}^{2}+{M_{{\rm Pl}}}^{2}FG_{,\phi X}-3\,{M_{{\rm Pl}}}^{2}F_{,\phi}G_{,X}\right)\dot{\phi}^{3}\nonumber \\
 &  & {}+[-112\,{M_{{\rm Pl}}}^{2}F{H}^{2}\xi_{,\phi}G_{,X}+16\,{M_{{\rm Pl}}}^{2}\xi_{,\phi}F_{,\phi}P_{,X}+8\,{M_{{\rm Pl}}}^{4}{F_{,\phi}}^{2}\xi_{,\phi\phi}-2\,{M_{{\rm Pl}}}^{4}FF_{,\phi}G_{,X}\nonumber \\
 &  & {}-256\,{M_{{\rm Pl}}}^{2}\xi_{,\phi}F_{,\phi}{H}^{2}\xi_{,\phi\phi}+16\,{M_{{\rm Pl}}}^{4}\xi_{,\phi}F_{,\phi}F_{,\phi\phi}-128\,{M_{{\rm Pl}}}^{2}{H}^{2}{\xi_{,\phi}}^{2}F_{,\phi\phi}+2\,{M_{{\rm Pl}}}^{4}{F}^{2}G_{,\phi X}\nonumber \\
 &  & {}-32\,{M_{{\rm Pl}}}^{2}\xi_{,\phi}F_{,\phi}G_{,\phi}+1536\,{H}^{4}{\xi_{,\phi}}^{2}\xi_{,\phi\phi}]\,\dot{\phi}^{2}\nonumber \\
 &  & {}+[8\, H\,(-384\,{\xi_{,\phi}}^{3}{H}^{4}+96\,{M_{{\rm Pl}}}^{2}{\xi_{,\phi}}^{2}F_{,\phi}{H}^{2}+8\,{M_{{\rm Pl}}}^{2}F\xi_{,\phi}G_{,\phi}-6\,{M_{{\rm Pl}}}^{4}{F_{,\phi}}^{2}\xi_{,\phi}\nonumber \\
 &  & {}+{M_{{\rm Pl}}}^{4}{F}^{2}G_{,X}-4\,{M_{{\rm Pl}}}^{2}FP_{,X}\xi_{,\phi})]\,\dot{\phi}\nonumber \\
 &  & {}+{M_{{\rm Pl}}}^{2}F\left(2\,{M_{{\rm Pl}}}^{2}FP_{,X}-4\,{M_{{\rm Pl}}}^{2}FG_{,\phi}+192\,{H}^{4}{\xi_{,\phi}}^{2}+3\,{M_{{\rm Pl}}}^{4}{F_{,\phi}}^{2}-48\,{M_{{\rm Pl}}}^{2}F_{,\phi}{H}^{2}\xi_{,\phi}\right).
\end{eqnarray}
\section{Third-order action}\label{app:intbp}

In the Appendices \ref{app:intbp}, \ref{app:intf3}, \ref{app:intf2}, 
and \ref{app:intf1}
we shall use the symbol $\doteq$
for the quantities which are valid up to total derivatives.

We first perturb the action $S=\int d^4 x\,\sqrt{-g}\,{\cal L}$
given in Eq.~(\ref{action}) up to third order. 
Since we choose the gauge in which the scalar field $\phi$ is 
unperturbed ($\delta \phi=0$), the functions $P$ and $G$ 
are expanded due to their $X$ dependence. 
After the Taylor-expansion of the action up to third order,
we can perform the following steps in order to simplify the result.
\begin{enumerate}
\item 
We start by removing the cubic term in $\psi$. 
One can employ the following relations 
\begin{eqnarray}
\label{B2}
c(t, {\bm x})\,\partial_{\hat{\imath}}^{3}\psi
&\doteq& -(\partial_{\hat{\imath}}c)(\partial_{\hat{\imath}}^{2}\psi)\qquad
{\rm \ for\ }\quad \hat{\imath}=1,2,3\,,\\
\label{B3}
c(t, {\bm x})\,(\partial_{\hat{\imath}\hat{\jmath}}\psi)
(\partial_{\hat{\jmath}}\partial_{\hat{\imath}}^{2}\psi)
&\doteq&-(\partial_{\hat{\imath}}c)\,(\partial_{\hat{\imath}\hat{\jmath}}\psi)\,
(\partial_{\hat{\imath}\hat{\jmath}}\psi)/2 \qquad
{\rm \ for\ }\quad \hat{\imath} \neq \hat{\jmath}\,,
\end{eqnarray}
where $c$ depends on other perturbation variables, and a hatted index
is not summed.
The $\psi$-cubic term automatically cancels out
after the integration by parts.

\item 
Now we simplify the term cubic in $\alpha$. 
This can be done by employing the following relations 
\begin{equation}
c(t)\alpha^{2}\dot{\alpha}\doteq-\dot{c}\,\alpha^{3}/3\,,
\qquad 
c(t,{\bm x})\,\partial_{\hat{\imath}}^{2}\alpha
\doteq-(\partial_{\hat{\imath}}\alpha)(\partial_{\hat{\imath}}c)\,.
\end{equation}

\item 
Next, we simplify the cubic term in ${\cal R}$. 
In this case we can make use
of the following relations
\begin{equation}
c(t)\dot{{\cal R}}^{2}\ddot{{\cal R}}\doteq -\dot{c}\,\dot{{\cal R}}^{3}/3\,,\quad
c(t){\cal R}^{2}\ddot{{\cal R}}\doteq \ddot{c}\,{\cal R}^{3}/3-2c{\cal R}\dot{{\cal R}}^{2}\,,
\quad c(t,{\bm x})\dot{{\cal R}}\ddot{{\cal R}}\doteq-\dot{c}\,\dot{{\cal R}}^{2}/2\,,
\quad c(t){\cal R}^{2}\dot{{\cal R}}\doteq-\dot{c}\,{\cal R}^{3}/3\,.
\end{equation}
Other useful relations are
\begin{eqnarray}
& & c(t)\ddot{{\cal R}}(\partial_{\hat{\imath}}{\cal R})^{2}\doteq  -\dot{{\cal R}}\,[\dot{c}\,(\partial_{\hat{\imath}}{\cal R})^{2}+2c\partial_{\hat{\imath}}{\cal R}\partial_{\hat{\imath}}\dot{{\cal R}}]\,,
\qquad c(t)\dot{{\cal R}}^{2}\partial_{\hat{\imath}}^{2}{\cal R}\doteq -2c\dot{{\cal R}}(\partial_{\hat{\imath}}{\cal R})(\partial_{\hat{\imath}}\dot{{\cal R}})\,,\nonumber \\
& & c(t){\cal R}\ddot{{\cal R}}\partial_{\hat{\imath}}^{2}{\cal R}\doteq c\,{\cal R}(\partial_{\hat{\imath}}\dot{{\cal R}})^{2}+\dot{c}\,[\dot{{\cal R}}(\partial_{\hat{\imath}}{\cal R})^{2}+{\cal R}(\partial_{\hat{\imath}}{\cal R})(\partial_{\hat{\imath}}\dot{{\cal R}})]+3c\,\dot{{\cal R}}(\partial_{\hat{\imath}}{\cal R})(\partial_{\hat{\imath}}\dot{{\cal R}})\,,\nonumber \\
& & c(t, {\bm x})\,\partial_{\hat{\imath}}^{2}{\cal R}\doteq -(\partial_{\hat{\imath}}c)(\partial_{\hat{\imath}}{\cal R})\,,\qquad c(t, {\bm x})(\partial_{\hat{\imath}}{\cal R})(\partial_{\hat{\imath}}\dot{{\cal R}})\doteq -\dot{c}\,(\partial_{\hat{\imath}}{\cal R})^{2}/2\,.
\end{eqnarray}
After performing these integrations by parts, we find that the ${\cal R}$-cubic
term can be written in the form 
$c_{1}(t){\cal R}^{3}+c_{2}(t){\cal R}\dot{{\cal R}}^{2}
+c_{3}(t)\dot{{\cal R}}^{3}+c_{4}(t){\cal R}\,(\partial_{i}{\cal R})^{2}$.

\item 
Now let us simplify the term quadratic in $\psi$ and linear in ${\cal R}$.
First we integrate by parts any derivative for the field ${\cal R}$,
so that these terms can be written as ${\cal R}\times({\rm quadratic\ term\ in\ }\psi)$.
Afterwards, we use the following relations (where $\hat{\imath} \neq \hat{\jmath}$)
\begin{eqnarray}
& & c(t, {\bm x})(\partial_{\hat{\imath}}^{2}\psi)(\partial_{\hat{\jmath}}^{2}\dot{\psi})\doteq  
-c(\partial_{\hat{\imath}}^{2}\dot{\psi})(\partial_{\hat{\jmath}}^{2}\psi)-\dot{c}(\partial_{\hat{\imath}}^{2}\psi)(\partial_{\hat{\jmath}}^{2}\psi)\,,\qquad c(t, {\bm x})(\partial_{\hat{\imath}\hat{\jmath}}\psi)(\partial_{\hat{\imath}\hat{\jmath}}\dot{\psi})\doteq -\dot{c}\,(\partial_{\hat{\imath}\hat{\jmath}}\psi)(\partial_{\hat{\imath}\hat{\jmath}}\psi)/2\,, \nonumber \\
& & c(t, {\bm x})\partial_{\hat{\imath}}^{3}\psi \doteq  -(\partial_{\hat{\imath}}c)(\partial_{\hat{\imath}}^{2}\psi)\,,\qquad c(t, {\bm x})(\partial_{\hat{\jmath}}\partial_{\hat{\imath}}^{2}\psi)\doteq -(\partial_{\hat{\imath}}^{2}\psi)(\partial_{\hat{\jmath}}c)\,.
\end{eqnarray}

\item 
Along the same lines we can simplify the term quadratic in $\psi$
and linear in $\alpha$ after integrating by parts any derivative 
of $\alpha$.
\item 
Let us next consider the term quadratic in $\alpha$ and linear in
${\cal R}$. In this case one can first integrate by parts 
all the second derivatives for the field $\alpha$. 
Then one can integrate the following first derivative terms
\begin{equation}
c(t, {\bm x})\alpha\dot{\alpha}\doteq -\dot{c}\,\alpha^{2}/2\,,
\qquad 
c(t, {\bm x})\alpha\partial_{\hat{\imath}}\alpha\doteq 
-\alpha^{2}\partial_{\hat{\imath}}c/2\,.
\end{equation}
\item 
Now we consider the term quadratic in ${\cal R}$. First of all
we can remove any derivative from the subset of terms which possess
the field $\alpha$. Then we eliminate the second derivatives
for the field ${\cal R}$ by using
\begin{eqnarray}
& & c(t, {\bm x})\dot{{\cal R}}\ddot{{\cal R}}\doteq  -\dot{c}\,\dot{{\cal R}}^{2}/2\,
\qquad c(t, {\bm x}){\cal R}\ddot{{\cal R}}\doteq -c\,\dot{{\cal R}}^{2}
+\ddot{c}\,{\cal R}^{2}/2\,,
\qquad c(t, {\bm x})\ddot{{\cal R}}\doteq -\dot{c}\dot{{\cal R}}\,, \nonumber \\
& & c(t)(\partial_{\hat{\imath}}\dot{\psi})(\partial_{\hat{\imath}}{\cal R})(\partial_{\hat{\imath}}^{2}{\cal R}) \doteq  -c\,(\partial_{\hat{\imath}}^{2}\dot{\psi})(\partial_{\hat{\imath}}{\cal R})^{2}/2\,,
\qquad c(t)(\partial_{\hat{\imath}\hat{\jmath}}\dot{\psi})(\partial_{\hat{\imath}}{\cal R})(\partial_{\hat{\jmath}}{\cal R})\doteq -c{\cal R}(\partial_{\hat{\imath}\hat{\jmath}}{\cal R})(\partial_{\hat{\imath}\hat{\jmath}}\dot{\psi})+c\,{\cal R}^{2}(\partial_{\hat{\imath}}^{2}\partial_{\hat{\jmath}}^{2}\dot{\psi})/2\,,\nonumber \\
& & c(t){\cal R}(\partial_{\hat{\imath}}^{2}{\cal R})(\partial_{\hat{\jmath}}^{2}\dot{\psi}) \doteq  -c(\partial_{\hat{\imath}}{\cal R})^{2}(\partial_{\hat{\jmath}}^{2}\dot{\psi})+c\,{\cal R}^{2}(\partial_{\hat{\imath}}^{2}\partial_{\hat{\jmath}}^{2}\dot{\psi})/2\,,\qquad c(t)(\partial_{\hat{\imath}\hat{\jmath}}{\cal R})(\partial_{\hat{\jmath}}{\cal R})(\partial_{\hat{\imath}}\dot{\psi})\doteq -c\,(\partial_{\hat{\jmath}}{\cal R})^{2}(\partial_{\hat{\imath}}^{2}\dot{\psi})/2\,,
\end{eqnarray}
where $\hat{\imath} \neq \hat{\jmath}$. 
The remaining terms can be simplified by using the following expressions
\begin{eqnarray}
& & c(t, {\bm x})\dot{{\cal R}}(\partial_{\hat{\imath}}\dot{{\cal R}}) \doteq   -\dot{{\cal R}}^{2}(\partial_{\hat{\imath}}c)/2\,,\qquad c(t, {\bm x}){\cal R}(\partial_{\hat{\imath}}{\cal R})\doteq -{\cal R}^{2}(\partial_{\hat{\imath}}c)/2\,,\qquad c(t, {\bm x}){\cal R}\dot{{\cal R}}\doteq -\dot{c}\,{\cal R}^{2}/2\,,
\nonumber \\
& & c(t)\dot{{\cal R}}(\partial_{\hat{\imath}}{\cal R})(\partial_{\hat{\imath}}\psi)\doteq 
\dot{c}\,{\cal R}^{2}\partial_{\hat{\imath}}^{2}\psi/2+c\,{\cal R}^{2}\partial_{\hat{\imath}}^{2}\dot{\psi}/2-c{\cal R}(\partial_{\hat{\imath}} \dot{\cal R})(\partial_{\hat{\imath}} \psi)\,,\qquad 
c(t, {\bm x})(\partial_{\hat{\imath}}\dot{{\cal R}})\doteq -\dot{{\cal R}}(\partial_{\hat{\imath}}c)\,.
\end{eqnarray}

\item 
For the terms which consist of a combination of three different
fields, we can integrate by parts any derivative of $\alpha$.
\end{enumerate}

\section{Integration of the term $f_{3}/a$}
\label{app:intf3}

We rewrite the integrand $f_3/a$ in Eq.~(\ref{S3simple}) in a more 
convenient form. We define
\begin{eqnarray}
\tilde X & = & c(t)\,{\cal R}\,[(\partial^{2}{\cal R})^{2}-(\partial_{i}\partial_{j}{\cal R})(\partial_{i}\partial_{j}{\cal R})]\,,\\
\tilde X_{k} & = & c(t)\,\dot{{\cal R}}\,[\partial_{k}{\cal R}\,(\partial^{2}{\cal R})-\partial_{i}{\cal R}(\partial_{i}\partial_{k}{\cal R})]-c(t)\,{\cal R}\,[(\partial_{k}\dot{{\cal R}})\,(\partial^{2}{\cal R})-(\partial_{i}\dot{{\cal R}})\,(\partial_{i}\partial_{k}{\cal R})]\,.
\end{eqnarray}
Then, for any function $c(t)$ dependent on $t$, we find
\begin{equation}
\dot{\tilde X}+2\partial_{k}\tilde X_{k}=
\dot{c}\,{\cal R}\,[(\partial^{2}{\cal R})^{2}
-(\partial_{i}\partial_{j}{\cal R})(\partial_{i}\partial_{j}{\cal R})]
+3c\,\dot{{\cal R}}\,[(\partial^{2}{\cal R})^{2}
-(\partial_{i}\partial_{j}{\cal R})
(\partial_{i}\partial_{j}{\cal R})]\,,
\end{equation}
which implies that 
\begin{equation}
c(t)\,\dot{{\cal R}}\,[(\partial^{2}{\cal R})^{2}-(\partial_{i}\partial_{j}{\cal R})
(\partial_{i}\partial_{j}{\cal R})]\doteq -\dot{c}\,{\cal R}\,[(\partial^{2}
{\cal R})^{2}-(\partial_{i}\partial_{j}{\cal R})(\partial_{i}\partial_{j}{\cal R})]/3\,.
\label{eq:tdw1}
\end{equation}

Let us also define
\begin{eqnarray}
Y & = & c(t)\,(\partial{\cal R})^{2}\,(\partial^{2}{\cal R})\,,\\
Y_{k} & = & c(t)\,[2{\cal R}\,\partial_{j}{\cal R}\partial_{k}
\partial_{j}{\cal R}-2{\cal R}\partial_{k}{\cal R}
\partial^{2}{\cal R}-\partial_{k}{\cal R}\,(\partial{\cal R})^{2}]\,.
\end{eqnarray}
It then follows that 
\begin{equation}
3Y+\partial_{k}Y_{k}=
-2c\,{\cal R}\,[(\partial^{2}{\cal R})^{2}
-(\partial_{i}\partial_{j}{\cal R})(\partial_{i}\partial_{j}{\cal R})]\,,
\end{equation}
so that
\begin{equation}
c(t)\,(\partial{\cal R})^{2}\,(\partial^{2}{\cal R})\doteq 
-2c\,{\cal R}\,[(\partial^{2}{\cal R})^{2}
-(\partial_{i}\partial_{j}{\cal R})(\partial_{i}\partial_{j}{\cal R})]/3\,.
\label{eq:tdw2}
\end{equation}
By using Eqs.~(\ref{eq:tdw1}) and (\ref{eq:tdw2}), one finds
\begin{eqnarray}
\frac{f_{3}}{a} & \doteq  & \frac{1}{a}\left[\left(A_{5}L_{1}^{2}-A_{7}L_{1}\right)\dot{{\cal R}}+A_{6}L_{1}^{2}{\cal R}\right][(\partial_{i}\partial_{j}{\cal R})(\partial_{i}\partial_{j}{\cal R})-(\partial^{2}{\cal R})^{2}]+\frac{A_{9}L_{1}^{2}}{a}\,(\partial{\cal R})^{2}\partial^{2}{\cal R}\nonumber \\
 & \doteq  & \frac{1}{a}\left[A_{6}L_{1}^{2}-\frac{1}{3}\,\frac{d}{dt}(A_{5}L_{1}^{2}-A_{7}L_{1})+\frac{H}{3}\,(A_{5}L_{1}^{2}-A_{7}L_{1})+\frac{2}{3}\, A_{9}L_{1}^{2}\right]{\cal R}\,[(\partial_{i}\partial_{j}{\cal R})(\partial_{i}\partial_{j}{\cal R})-(\partial^{2}{\cal R})^{2}]\,.
 \label{eq:tdf31}
\end{eqnarray}
Employing the following relation
\begin{eqnarray}
\hspace{-0.5cm}
c(t)\,{\cal R}\,[(\partial_{i}\partial_{j}{\cal R})(\partial_{i}\partial_{j}{\cal R})-(\partial^{2}{\cal R})^{2}] & \doteq  & -c\,(\partial_{i}{\cal R})\,(\partial_{j}{\cal R})(\partial_{i}\partial_{j}{\cal R})-c\,(\partial_{i}{\cal R})\,{\cal R}\,(\partial_{i}\partial^{2}{\cal R})-c\,{\cal R}\,(\partial^{2}{\cal R})^{2}
\nonumber \\
& \doteq  & -c\,{\cal R}\,\partial_{i}\partial_{j}[(\partial_{i}{\cal R})\,(\partial_{j}{\cal R})]+c\,(\partial{\cal R})^{2}\,(\partial^{2}{\cal R})\,,
\label{eq:Orr}
\end{eqnarray}
we obtain the form of $f_3/a$ given in Eq.~(\ref{eq:f3expr2}).

\section{Integration of the term $af_{2}$}
\label{app:intf2}

Regarding the integrands of the term $af_{2}$ in Eq.~(\ref{S3simple}), 
we start by considering the contribution written in the form
$c(t)\,\dot{{\cal R}}(\partial_{i}\partial_{j}{\cal R})(\partial_{i}\partial_{j}{\cal X})$.
We define
\begin{eqnarray}
Z &=& c(t)\,{\cal R}\,(\partial_{i}\partial_{j}{\cal R})
(\partial_{i}\partial_{j}{\cal X})\,,\\
Z_k &=& c(t)\,(\dot{{\cal R}}\partial_{i}{\cal R}-{\cal R}\partial_{i}\dot{{\cal R}})\,(\partial_{k}\partial_{i}{\cal X})\,,
\end{eqnarray}
so that
\begin{equation}
\dot{Z}+\partial_{k}Z_{k}=\dot{c}\,{\cal R}\,(\partial_{i}\partial_{j}{\cal R})(\partial_{i}\partial_{j}{\cal X})+2c\dot{{\cal R}}(\partial_{i}\partial_{j}{\cal R})(\partial_{i}\partial_{j}{\cal X})+c{\cal R}\,(\partial_{i}\partial_{j}{\cal R})(\partial_{i}\partial_{j}\dot{{\cal X}})+c\, Q\,(\dot{{\cal R}}\partial_{i}{\cal R}-{\cal R}\partial_{i}\dot{{\cal R}})\,\partial_{i}\dot{{\cal R}}\,.
\end{equation}
This gives
\begin{eqnarray}
c(t)\,\dot{{\cal R}}(\partial_{i}\partial_{j}{\cal R})(\partial_{i}\partial_{j}{\cal X}) & \doteq  & -[(\dot{c}+a q_2)\,{\cal R}\,(\partial_{i}\partial_{j}{\cal R})(\partial_{i}\partial_{j}{\cal X})+c\,{\cal R}\,(\partial_{i}\partial_{j}{\cal R})(\partial_{i}\partial_{j}\dot{{\cal X}})+c\, Q\,(\dot{{\cal R}}\partial_{i}{\cal R}-{\cal R}\partial_{i}
\dot{{\cal R}})\,\partial_{i}\dot{{\cal R}}]/2\nonumber \\
 &  &{} +aq_2\,{\cal R}\,(\partial_{i}\partial_{j}{\cal R})
 (\partial_{i}\partial_{j}{\cal X})/2\,.
\end{eqnarray}
For later convenience we added and subtracted the quantity 
$aq_2\,{\cal R}\,(\partial_{i}\partial_{j}{\cal R})(\partial_{i}\partial_{j}{\cal X})/2$.

Let us focus on the term containing $q_2$ alone. Then we have
\begin{eqnarray}
a q_2(t)\,{\cal R}\,[(\partial_{i}\partial_{j}{\cal R})(\partial_{i}\partial_{j}{\cal X})]/2 
& \doteq  & [-aq_2\,(\partial_{i}{\cal X})\,(\partial_{j}{\cal R})(\partial_{i}\partial_{j}{\cal R})-aq_2\,(\partial_{i}{\cal X})\,{\cal R}\,(\partial_{i}\partial^{2}{\cal R})]/2
\nonumber \\
& \doteq  & aq_2\,\{(\partial_{i}{\cal X})\,(\partial_{i}{\cal R})\,(\partial^{2}{\cal R})-{\cal R}\,\partial_{i}\partial_{j}[(\partial_{i}{\cal R})\,(\partial_{j}{\cal X})]\}/2
+aq_2\,Q\,{\cal R}\dot{{\cal R}}(\partial^{2}{\cal R})/2 \,.
\label{eq:Orr-1}
\end{eqnarray}

For the term 
$(\partial_{i}{\cal R})\,(\partial_{i}{\cal X})\,(\partial^{2}{\cal R})$
the following equalities hold
\begin{eqnarray}
c(t)\,(\partial_{i}{\cal R})\,(\partial_{i}{\cal X})\,(\partial^{2}{\cal R}) & \doteq  & -c\partial_{ik}{\cal R}\partial_{i}{\cal X}\partial_{k}{\cal R}-c\partial_{i}{\cal R}\partial_{ik}{\cal X}\partial_{k}{\cal R}\doteq 2c{\cal R}\partial_{ik}{\cal R}\partial_{ik}{\cal X}+c{\cal R}\partial_{i}\partial^{2}{\cal R}\partial_{i}{\cal X}+c{\cal R}\partial_{i}{\cal R}\partial_{i}\partial^{2}{\cal X} \nonumber \\
 & \doteq  & 2c{\cal R}\partial_{ik}{\cal R}\partial_{ik}{\cal X}-c\partial_{i}{\cal R}\partial_{i}{\cal X}\partial^{2}{\cal R}-c{\cal R}\partial^{2}{\cal R}\partial^{2}{\cal X}+c{\cal R}\partial_{i}{\cal R}\partial_{i}\partial^{2}{\cal X}\,,
\end{eqnarray}
so that
\begin{eqnarray}
c(t)\,(\partial_{i}{\cal R})\,(\partial_{i}{\cal X})\,(\partial^{2}{\cal R}) & \doteq  & c{\cal R}\partial_{ik}{\cal R}\partial_{ik}{\cal X}-cQ\,{\cal R}(\partial^{2}{\cal R})\dot{{\cal R}}/2+cQ\,{\cal R}(\partial_{i}{\cal R})(\partial_{i}\dot{{\cal R}})/2\nonumber \\
 & \doteq  & c\,{\cal R}\,(\partial_{i}\partial_{j}{\cal R})(\partial_{i}\partial_{j}{\cal X})+c\, Q\,{\cal R}(\partial_{i}{\cal R})(\partial_{i}\dot{{\cal R}})+cQ\,\dot{{\cal R}}\,(\partial{\cal R})^{2}/2\,,
\end{eqnarray}
where we have defined $\partial_{ij}A\equiv\partial_{i}\partial_{j}A$.
Putting all these partial results together, it follows that 
\begin{eqnarray}
& & a\left[-\frac{2A_{6}L_{1}}{w_{1}}\,{\cal R}(\partial_{i}\partial_{j}{\cal R})(\partial_{i}\partial_{j}{\cal X})+\frac{1}{w_{1}}\left(A_{7}-2A_{5}L_{1}\right)\dot{{\cal R}}(\partial_{i}\partial_{j}{\cal R})(\partial_{i}\partial_{j}{\cal X})-\frac{A_{9}L_{1}}{w_{1}}\partial^{2}{\cal R}\partial_{i}{\cal R}\partial_{i}{\cal X}
\right]\nonumber \\
&& \doteq -a\left[\frac{2A_{6}L_{1}}{w_{1}}+\frac{d}{dt}\!\left(\frac{A_{7}-2A_{5}L_{1}}{2w_{1}}\right)+H\left(\frac{A_{7}-2A_{5}L_{1}}{2w_{1}}\right)+\frac{q_2}{2}+\frac{A_{9}L_{1}}{w_{1}}\right]{\cal R}(\partial_{i}\partial_{j}{\cal R})(\partial_{i}\partial_{j}{\cal X})\nonumber \\
 &  &~~~\,-\frac{a}{2w_{1}}\,(A_{7}-2A_{5}L_{1})\,{\cal R}\,(\partial_{i}\partial_{j}{\cal R})(\partial_{i}\partial_{j}\dot{{\cal X}})+\frac{aq_2}{2}\,\{(\partial_{i}{\cal X})\,(\partial_{i}{\cal R})\,(\partial^{2}{\cal R})-{\cal R}\,\partial_{i}\partial_{j}[(\partial_{i}{\cal R})\,(\partial_{j}{\cal X})]\}\nonumber \\
 &  &~~~\,-\frac{aQ}{2w_{1}}\,(A_{7}-2A_{5}L_{1})\,(\dot{{\cal R}}\partial_{i}{\cal R}-{\cal R}\partial_{i}\dot{{\cal R}})\,\partial_{i}\dot{{\cal R}}+\frac{a q_2 Q}{2}\,{\cal R}\dot{{\cal R}}(\partial^{2}{\cal R})
 -\frac{a}{w_{1}}\, A_{9}L_{1}\, Q{\cal R}(\partial_{i}{\cal R})(\partial_{i}\dot{{\cal R}}) \nonumber \\
& &~~~\, -\frac{a}{2w_{1}}\, A_{9}L_{1}\, Q\dot{{\cal R}}\,(\partial{\cal R})^{2}\,.
\label{eq:f2aux1-2}
\end{eqnarray}
We choose $q_2$ such that 
\begin{equation}
\frac{2A_{6}L_{1}}{w_{1}}+\frac{d}{dt}\!\left(\frac{A_{7}-2A_{5}L_{1}}{2w_{1}}\right)+H\left(\frac{A_{7}-2A_{5}L_{1}}{2w_{1}}\right)+\frac{q_2}{2}+\frac{A_{9}L_{1}}{w_{1}}=\frac{3H}{2w_{1}}\,(A_{7}-2A_{5}L_{1})\,.
\label{q2def}
\end{equation}
In this case we have 
\begin{eqnarray}
\hspace{-0.4cm}
& & a\left[-\frac{2A_{6}L_{1}}{w_{1}}\,{\cal R}(\partial_{i}\partial_{j}{\cal R})(\partial_{i}\partial_{j}{\cal X})+\frac{1}{w_{1}}\left(A_{7}-2A_{5}L_{1}\right)\dot{{\cal R}}(\partial_{i}\partial_{j}{\cal R})(\partial_{i}\partial_{j}{\cal X})-\frac{A_{9}L_{1}}{w_{1}}\partial^{2}{\cal R}\partial_{i}{\cal R}\partial_{i}{\cal X}\right]
\nonumber \\
\hspace{-0.4cm}& &\doteq  -\frac{a}{2w_{1}}\,(A_{7}-2A_{5}L_{1})\,[{\cal R}\,(\partial_{i}\partial_{j}{\cal R})(\partial_{i}\partial_{j}\dot{{\cal X}})+3H{\cal R}(\partial_{i}\partial_{j}{\cal R})(\partial_{i}\partial_{j}{\cal X})]
+\frac{aq_2}{2}\,\{(\partial_{i}{\cal X})\,(\partial_{i}{\cal R})\,(\partial^{2}{\cal R})-{\cal R}\,\partial_{i}\partial_{j}[(\partial_{i}{\cal R})\,(\partial_{j}{\cal X})]\}\nonumber \\
\hspace{-0.4cm} &  &~~~\,-\frac{aQ}{2w_{1}}\,(A_{7}-2A_{5}L_{1})\,(\dot{{\cal R}}\partial_{i}{\cal R}-{\cal R}\partial_{i}\dot{{\cal R}})\,\partial_{i}\dot{{\cal R}}+\frac{aq_2Q}{2}\,{\cal R}\dot{{\cal R}}(\partial^{2}{\cal R}) -\frac{a}{w_{1}}\, A_{9}L_{1}\, Q{\cal R}(\partial_{i}{\cal R})(\partial_{i}\dot{{\cal R}}) \nonumber \\
\hspace{-0.4cm}& &~~~\,-\frac{a}{2w_{1}}\, A_{9}L_{1}\, Q\dot{{\cal R}}\,(\partial{\cal R})^{2}\,.
\label{eq:f2aux1}
\end{eqnarray}

Consider the following combination
\begin{eqnarray}
& & a c(t)\,{\cal R}(\partial_{i}\partial_{j}{\cal R})(\partial_{i}\partial_{j}\dot{{\cal X}})+3aH\, c(t)\,{\cal R}(\partial_{i}\partial_{j}{\cal R})(\partial_{i}\partial_{j}{\cal X}) \nonumber \\
& & \doteq -ac(\partial_{i}{\cal R})\,[(\partial_{j}{\cal R})(\partial_{i}\partial_{j}\dot{{\cal X}})+{\cal R}\partial_{i}\partial^{2}\dot{{\cal X}}]
-3aHc(\partial_{i}{\cal R})\,[(\partial_{j}{\cal R})(\partial_{i}\partial_{j}{\cal X})+{\cal R}\partial_{i}\partial^{2}{\cal X}]\nonumber \\
 &&\doteq  -ac\dot{{\cal X}}\partial_{i}\partial_{j}[(\partial_{i}{\cal R})(\partial_{j}{\cal R})]+ac\,(\partial^{2}\dot{{\cal X}})\,[(\partial{\cal R})^{2}+{\cal R}\partial^{2}{\cal R}]
 -3aHc{\cal X}\partial_{i}\partial_{j}[(\partial_{i}{\cal R})(\partial_{j}{\cal R})]
 +3aHc\,(\partial^{2}{\cal X})\,[(\partial{\cal R})^{2}+{\cal R}\partial^{2}{\cal R}]\nonumber \\
 & & \doteq   ac\,(\partial{\cal R})^{2}\,[\partial^{2}(\dot{{\cal X}}+3H{\cal X})]-ac\,(\dot{{\cal X}}+3H{\cal X})\partial_{i}\partial_{j}[(\partial_{i}{\cal R})(\partial_{j}{\cal R})]+ac\,{\cal R}\partial^{2}{\cal R}\,(\partial^{2}\dot{{\cal X}})+3aH\, Q\, c\,{\cal R}\,\dot{{\cal R}}\partial^{2}{\cal R}\nonumber \\
 & & \doteq \frac{c}{a^{2}}\,\frac{d}{dt}(a^{3}Q\,\dot{{\cal R}})\,\{(\partial{\cal R})^{2}-\partial^{-2}\partial_{i}\partial_{j}[(\partial_{i}{\cal R})(\partial_{j}{\cal R})]\}+ac\,{\cal R}\partial^{2}{\cal R}\,(\partial^{2}\dot{{\cal X}})+3aH\, Q\, c\,{\cal R}\,\dot{{\cal R}}\partial^{2}{\cal R}\,,
\end{eqnarray}
where we have used the properties
\begin{equation}
{\cal X}=Q\partial^{-2}\dot{{\cal R}}\,,\qquad{\rm and\qquad}
\partial^{2}(\dot{{\cal X}}+3H{\cal X})=\left(\frac{d}{dt}+3H\right)
(Q\dot{{\cal R}})=a^{-3}\frac{d}{dt}(a^{3}Q\,\dot{{\cal R}})\,.
\label{eq:nablCC}
\end{equation}
The reason why we have introduced this combination is that it is a part of
the linear equation of motion (\ref{linearR2}).
Then Eq.~(\ref{eq:f2aux1}) reduces to
\begin{eqnarray}
\hspace{-0.5cm}& & a\left[-\frac{2A_{6}L_{1}}{w_{1}}\,{\cal R}(\partial_{i}\partial_{j}{\cal R})(\partial_{i}\partial_{j}{\cal X})+\frac{1}{w_{1}}\left(A_{7}-2A_{5}L_{1}\right)\dot{{\cal R}}(\partial_{i}\partial_{j}{\cal R})(\partial_{i}\partial_{j}{\cal X})-\frac{A_{9}L_{1}}{w_{1}}\partial^{2}{\cal R}\partial_{i}{\cal R}\partial_{i}{\cal X}\right]\nonumber \\
\hspace{-0.5cm} & &\doteq -\frac{A_{7}-2A_{5}L_{1}}{2w_{1}a^{2}} \frac{d}{dt}(a^{3}Q\,\dot{{\cal R}})\,\{(\partial{\cal R})^{2}-\partial^{-2}\partial_{i}\partial_{j}[(\partial_{i}{\cal R})(\partial_{j}{\cal R})]\}
 +\frac{aq_2}{2}\{(\partial_{i}{\cal X})\,(\partial_{i}{\cal R})\,(\partial^{2}{\cal R})-{\cal R}\,\partial_{i}\partial_{j}[(\partial_{i}{\cal R})\,(\partial_{j}{\cal X})]\}\nonumber \\
\hspace{-0.5cm} & &~~~\,-\frac{a}{2w_{1}}\,(A_{7}-2A_{5}L_{1})\,{\cal R}\partial^{2}{\cal R}\,(\partial^{2}\dot{{\cal X}})-\frac{3aH}{2w_{1}}\, Q\,(A_{7}-2A_{5}L_{1})\,{\cal R}\,\dot{{\cal R}}\partial^{2}{\cal R}-\frac{aQ}{2w_{1}}\,(A_{7}-2A_{5}L_{1})\,(\dot{{\cal R}}\partial_{i}{\cal R}-{\cal R}\partial_{i}\dot{{\cal R}})\,\partial_{i}\dot{{\cal R}} \nonumber \\
\hspace{-0.5cm}& &~~~\,+\frac{aq_2Q}{2}\,{\cal R}\dot{{\cal R}}(\partial^{2}{\cal R})
-\frac{a}{w_{1}}\, A_{9}L_{1}\, Q{\cal R}(\partial_{i}{\cal R})(\partial_{i}\dot{{\cal R}})-\frac{a}{2w_{1}}\, A_{9}L_{1}\, Q\dot{{\cal R}}\,(\partial{\cal R})^{2}\,.
\label{eq:f2aux1-1}
\end{eqnarray}
Combining these terms with the other ones in the $af_{2}$ term,
we find 
\begin{eqnarray}
af_{2} & \doteq  & -\frac{A_{7}-2A_{5}L_{1}}{2w_{1}a^{2}} \frac{d}{dt}(a^{3}Q\,\dot{{\cal R}})\,\{(\partial{\cal R})^{2}-\partial^{-2}\partial_{i}\partial_{j}[(\partial_{i}{\cal R})(\partial_{j}{\cal R})]\}
+\frac{aq_2}{2}\,\{(\partial_{i}{\cal X})\,(\partial_{i}{\cal R})\,(\partial^{2}{\cal R})-{\cal R}\,\partial_{i}\partial_{j}[(\partial_{i}{\cal R})\,(\partial_{j}{\cal X})]\}\nonumber \\
 &  &{} +a\left[A_{6}\frac{2L_{1}Q}{w_{1}}-\frac{3H}{2w_{1}}\, Q\,(A_{7}-2A_{5}L_{1})
 +\frac{q_2 Q}{2}\right]{\cal R}\dot{{\cal R}}\partial^{2}{\cal R}
-\frac{3a}{2w_{1}}\, A_{9}L_{1}\, Q \dot{{\cal R}}(\partial{\cal R})^{2}
 +a\, A_{8}{\cal R}(\partial{\cal R})^{2}\nonumber \\
 &  &{} +a\left(A_{2}-A_{3}L_{1}+A_{5}\frac{2L_{1}Q}{w_{1}}-A_{7}\frac{Q}{w_{1}}\right)\dot{{\cal R}}^{2}\partial^{2}{\cal R}-\frac{a}{2w_{1}}\,(A_{7}-2A_{5}L_{1})\,{\cal R}\,(\partial^{2}{\cal R})\,(\partial^{2}\dot{{\cal X}})\nonumber \\
 &  &{} -\frac{aQ}{2w_{1}}\,(A_{7}-2A_{5}L_{1})\,(\dot{{\cal R}}\partial_{i}{\cal R}-{\cal R}\partial_{i}\dot{{\cal R}})\,\partial_{i}\dot{{\cal R}}-\frac{a}{w_{1}}\, A_{9}L_{1}\, Q{\cal R}(\partial_{i}{\cal R})(\partial_{i}\dot{{\cal R}}).
\label{eq:af2-1}
\end{eqnarray}
This is similar to the formula (131) of Ref.~\cite{Hael}. 
Let us define
\begin{align}
{\cal A}& =  {\cal R}\dot{{\cal R}}\partial^{2}{\cal R}\,,& {\cal B}& =\dot{{\cal R}}(\partial{\cal R})^{2}\,,
&{\cal C} & = {\cal R}\,(\partial{\cal R})^{2}\,,&{\cal D} & = \dot{{\cal R}}^{2}(\partial^{2}{\cal R})\,,&
{\cal E} &= {\cal R}\,(\partial^{2}{\cal R})(\partial^{2}\dot{{\cal X}}) \,,\nonumber \\
{\cal F}&=\dot{{\cal R}}\partial_{i}{\cal R}\partial_{i}\dot{{\cal R}}  \,,
&{\cal H} & = {\cal R}\partial_{i}\dot{{\cal R}}\partial_{i}\dot{{\cal R}} \,,
&{\cal I} & ={\cal R}\partial_{i}{\cal R}\partial_{i}\dot{{\cal R}}\,,
&{\cal J} & ={\cal R}^{2}\partial^{2}{\cal R}\,,
&{\cal K}& ={\cal R}^{2}\partial^{2}\dot{{\cal R}}\,.
\end{align}
Then the following relations hold 
\begin{eqnarray}
{\cal B} & \doteq &  -{\cal A}-{\cal I}\,,\qquad {\cal D}  \doteq   -2{\cal F}\,,\qquad 
{\cal K}  \doteq   -2{\cal I}\,,\qquad {\cal J}  \doteq   -2{\cal C}\,,
\qquad c{\cal B}  \doteq   -\dot{c}\,{\cal C}-2c{\cal I}\,,\nonumber \\
c\,{\cal E} & \doteq &  c{\cal R}\,(\partial^{2}{\cal R})(\dot{Q}\dot{{\cal R}}+Q\ddot{{\cal R}})\doteq c\,\dot{Q} {\cal A}-\dot{{\cal R}}\,[(\dot{c}\,Q+c\,\dot{Q}){\cal R}\partial^{2}{\cal R}+c\,Q\dot{{\cal R}}\partial^{2}{\cal R}+c\,Q{\cal R}\partial^{2}\dot{{\cal R}}]
\nonumber \\
 & \doteq  & -\dot{c}\,Q{\cal A}-c\,Q{\cal D}+c\,Q\partial_{i}\dot{{\cal R}}\,({\cal R}\partial_{i}\dot{{\cal R}}+\dot{{\cal R}}\partial_{i}{\cal R})\doteq Q 
 (-\dot{c}\,{\cal A}-c\,{\cal D}+c{\cal H}+c{\cal F})\,.
\end{eqnarray}
Because of these identities, we can rewrite Eq.~(\ref{eq:af2-1})
in the form
\begin{eqnarray}
af_{2} & \doteq  & -\frac{A_{7}-2A_{5}L_{1}}{2w_{1}a^{2}}\frac{d}{dt}(a^{3}Q\,\dot{{\cal R}})\,\{(\partial{\cal R})^{2}-\partial^{-2}\partial_{i}\partial_{j}[(\partial_{i}{\cal R})(\partial_{j}{\cal R})]\}
+\frac{aq_2}{2}\,\{(\partial_{i}{\cal X})\,(\partial_{i}{\cal R})\,(\partial^{2}{\cal R})-{\cal R}\,\partial_{i}\partial_{j}[(\partial_{i}{\cal R})\,(\partial_{j}{\cal X})]\}\nonumber \\
 &  &{} +a\left[A_{6}\frac{2L_{1}Q}{w_{1}}-\frac{3H}{2w_{1}}\, Q\,(A_{7}-2A_{5}L_{1})+\frac{q_2Q}{2}+Q\,\frac{d}{dt}\!\left(\frac{A_{7}-2A_{5}L_{1}}{2w_{1}}\right)+\frac{QH}{2w_{1}}\,(A_{7}-2A_{5}L_{1})\right]{\cal R}\dot{{\cal R}}\partial^{2}{\cal R}\nonumber \\
&  &{} -\frac{3aA_9}{2} \frac{L_{1}Q}{w_1}
\dot{{\cal R}}(\partial{\cal R})^{2}+a\, A_{8}{\cal R}(\partial{\cal R})^{2}
+a\left(A_{2}-A_{3}L_{1}\right)\dot{{\cal R}}^{2}\partial^{2}{\cal R}
-aA_9 \frac{L_1Q}{w_1}
{\cal R}(\partial_{i}{\cal R})(\partial_{i}\dot{{\cal R}}) \,.
\label{eq:af2-1-1}
\end{eqnarray}
Using the relations $A_9=-2w_1$, 
$c\,{\cal I}  \doteq  -\dot{c}\,{\cal C}+c\,{\cal A},~c\,{\cal B}
\doteq \dot{c}\,{\cal C}-2c\,{\cal A}$ and the definition of 
$q_2$, we obtain the form of $af_2$ given in Eq.~(\ref{eq:f2expr2}).

\section{Integration of the term $a^3 f_{1}$}
\label{app:intf1}

By defining
\begin{eqnarray}
W &= & c(t)\,{\cal R}\,(\partial_{kl}{\cal X})(\partial_{kl}{\cal X})\,,\\
W_{k} &=& c(t)\,{\cal R}\,(\partial_{j}{\cal X})(\partial_{kj}\dot{{\cal X}})
-c(t)\partial_{k}{\cal R}(\partial_{j}{\cal X})(\partial_{j}\dot{{\cal X}})
+c(t)\,\dot{{\cal X}}\partial_{j}[(\partial_{j}{\cal R})(\partial_{k}{\cal X})]\,,
\end{eqnarray}
we have 
\begin{equation}
2\partial_{k}W_{k}-\dot{W}-2c{\cal R}\partial_{k}{\cal X}\partial_{k}\partial^{2}\dot{{\cal X}}-2c\dot{{\cal X}}\partial_{kl}(\partial_{k}{\cal R}\partial_{l}{\cal X})+c\dot{{\cal R}}(\partial_{ik}{\cal X})(\partial_{ik}{\cal X})=-\dot{c}{\cal R}(\partial_{ik}{\cal X})(\partial_{ik}{\cal X})\,.
\end{equation}
It follows that 
\begin{eqnarray}
c(t)\,\dot{{\cal R}}(\partial_{ik}{\cal X})(\partial_{ik}{\cal X}) & \doteq  & -\dot{c}{\cal R}(\partial_{ik}{\cal X})(\partial_{ik}{\cal X})+2c{\cal R}(\partial_{k}{\cal X})(\partial_{k}\partial^{2}\dot{{\cal X}})+2c\dot{{\cal X}}\partial_{kl}(\partial_{k}{\cal R}\partial_{l}{\cal X})\nonumber \\
 & \doteq  & -\dot{c}{\cal R}(\partial_{ik}{\cal X})(\partial_{ik}{\cal X})+2c\dot{{\cal X}}\partial_{kl}(\partial_{k}{\cal R}\partial_{l}{\cal X})-2c(\partial_{k}{\cal R})(\partial_{k}{\cal X})\partial^{2}\dot{{\cal X}}-2c{\cal R}(\partial^{2}{\cal X})(\partial^{2}\dot{{\cal X}})\,.
\label{eq:f1aux1-1}
\end{eqnarray}
There is also another relation
\begin{eqnarray}
d(t)\,{\cal R}(\partial_{kl}{\cal X})(\partial_{kl}{\cal X}) & \doteq  & 
-d(\partial_{k}{\cal X})(\partial_{l}{\cal R})(\partial_{kl}{\cal X})
-d(\partial_{k}{\cal X}){\cal R}(\partial_{k}\partial^{2}{\cal X})\nonumber \\
& \doteq  & -d{\cal X}\,\partial_{kl}[(\partial_{k}{\cal X})(\partial_{l}{\cal R})]
+d(\partial_{k}{\cal X})(\partial_{k}{\cal R})(\partial^{2}{\cal X})
+d{\cal R}\,(\partial^{2}{\cal X})(\partial^{2}{\cal X})\,.
\label{eq:f1aux1-2}
\end{eqnarray}

We study the contribution of the last term in Eq.~(\ref{eq:f1expr}), i.e.
$(c_1 \dot{\cal R}+c_2 {\cal R})(\partial_i \partial_j {\cal X})(\partial_i \partial_j {\cal X})$, 
where $c_{1}=A_{5}/w_{1}^{2}$ and $c_{2}=A_{6}/w_{1}^{2}$.
Using Eqs.~(\ref{eq:f1aux1-1}) and (\ref{eq:f1aux1-2}), we have
\begin{eqnarray}
& & a^{3} c_{1}\dot{{\cal R}}(\partial_{ik}{\cal X})(\partial_{ik}{\cal X})+a^{3}\,(c_{2}-p_{1})\,{\cal R}(\partial_{kl}{\cal X})(\partial_{kl}{\cal X})+a^{3}\,p_1{\cal R}
(\partial_{kl}{\cal X})(\partial_{kl}{\cal X}) \nonumber \\
& & \doteq -a^{3}(\dot{c}_{1}+3Hc_{1}-p_{1}){\cal R}(\partial_{ik}{\cal X})(\partial_{ik}{\cal X})
+a^{3}[2c_{1}\dot{{\cal X}}\partial_{kl}(\partial_{k}{\cal R}\partial_{l}{\cal X})-2c_{1}\partial_{k}{\cal R}\partial_{k}{\cal X}\partial^{2}\dot{{\cal X}}-2c_{1}{\cal R}(\partial^{2}{\cal X})
(\partial^{2}\dot{{\cal X}})] \nonumber \\
& &~~~+a^{3}[(p_{1}-c_{2}){\cal X}\,\partial_{kl}[(\partial_{k}{\cal X})(\partial_{l}{\cal R})]
+(c_{2}-p_{1})(\partial_{k}{\cal X})(\partial_{k}{\cal R})(\partial^{2}{\cal X})
+(c_{2}-p_{1}){\cal R}\,(\partial^{2}{\cal X})(\partial^{2}{\cal X})] \,,
\end{eqnarray}
where we added and subtracted the term 
$a^{3} p_{1}{\cal R}(\partial_{kl}{\cal X})(\partial_{kl}{\cal X})$.
Let us define $p_{1}$ such that
\begin{equation}
p_{1}-c_{2}=6Hc_{1}\,,
\end{equation}
in which case
\begin{eqnarray}
& & a^{3} c_{1}\dot{{\cal R}}(\partial_{ik}{\cal X})(\partial_{ik}{\cal X}) 
+ a^{3}\,(c_{2}-p_{1})\,{\cal R}(\partial_{kl}{\cal X})(\partial_{kl}{\cal X})
+a^{3}\, p_{1}{\cal R}(\partial_{kl}{\cal X})
(\partial_{kl}{\cal X}) \nonumber \\
& &\doteq -a^{3}(\dot{c}_{1}+3Hc_{1}-p_{1})
{\cal R}(\partial_{ik}{\cal X})(\partial_{ik}{\cal X}) \nonumber \\
& &~~~+2c_{1}a^{3}\{(\dot{{\cal X}}+3H{\cal X})\partial_{kl}(\partial_{k}{\cal R}\partial_{l}{\cal X})-(\partial_{k}{\cal R})(\partial_{k}{\cal X})\partial^{2}(\dot{{\cal X}}+3H{\cal X})-{\cal R}(\partial^{2}{\cal X})[\partial^{2}(\dot{{\cal X}}+3H{\cal X})]\}\nonumber \\
& & \doteq -a^{3}(\dot{c}_{1}+3Hc_{1}-p_{1})
{\cal R}(\partial_{ik}{\cal X})(\partial_{ik}{\cal X})-2c_{1}\frac{d}{dt}(a^{3}Q\,\dot{{\cal R}})\,\{(\partial_{k}{\cal R})(\partial_{k}{\cal X})-\partial^{-2}\partial_{i}\partial_{j}[(\partial_{i}{\cal R})(\partial_{j}{\cal X})]\}\nonumber \\
 &  &~~~-2c_{1}\, Q\,{\cal R}\,\dot{{\cal R}}\,\frac{d}{dt}(a^{3}Q\,\dot{{\cal R}})\,,
 \label{eq:f1aux1}
\end{eqnarray}
where we used Eq.~(\ref{eq:nablCC}). 
The last term of Eq.~(\ref{eq:f1aux1})
is expressed as
\begin{eqnarray}
\hspace{-0.5cm}
-2c_{1}\, Q\,{\cal R}\,\dot{{\cal R}}\,\frac{d}{dt}(a^{3}Q\,\dot{{\cal R}})\doteq  
-\frac{c_{1}}{a^{3}}\,{\cal R}\,\frac{d}{dt}[(a^{3}Q\,\dot{{\cal R}})^{2}]
&\doteq & (a^{3}Q\,\dot{{\cal R}})^{2}\,\frac{d}{dt}\left(\frac{c_{1}{\cal R}}{a^{3}}\right)
\nonumber \\
\hspace{-0.5cm}
&\doteq & a^{3}c_{1}\, Q^{2}\dot{{\cal R}}^{3}+a^{3}\,(\dot{c}_{1}-3Hc_{1})
Q^{2}\,{\cal R}\dot{{\cal R}}{}^{2}\,.
\end{eqnarray}

The integrand $a^3f_1$ can be written as
\begin{eqnarray}
a^{3}f_{1} & \doteq  & a^{3}\left(A_{1}+A_{3}\frac{Q}{w_{1}}-q_{1}\, Q\right)\dot{{\cal R}}^{3}+a^{3}\left[A_{4}-A_{6}\frac{Q^{2}}{w_{1}^{2}}+Q^{2}\,\frac{d}{dt}\left(\frac{A_{5}}{w_{1}^{2}}\right)-\frac{3HA_{5}Q^{2}}{w_{1}^{2}}\right]{\cal R}\dot{{\cal R}}^{2}
+a^{3}q_{1}Q\dot{{\cal R}}^{3}\nonumber \\
 &  &{} +a^{3}\left[\frac{A_{6}}{w_{1}^{2}}-\frac{d}{dt}\left(\frac{A_{5}}{w_{1}^{2}}\right)+\frac{3HA_{5}}{w_{1}^{2}}\right]{\cal R}(\partial_{ik}{\cal X})(\partial_{ik}{\cal X})-\frac{2A_{5}}{w_{1}^{2}}\,\frac{d}{dt}(a^{3}Q\,\dot{{\cal R}})[(\partial_{k}{\cal R})(\partial_{k}{\cal X})-\partial^{-2}\partial_{i}\partial_{j}[(\partial_{i}{\cal R})(\partial_{j}{\cal X})]]\nonumber \\
 &  &{} +\frac{a^{3}}{w_{1}}\left(A_{9}\frac{Q}{w_{1}}\right)\dot{{\cal R}}\partial_{i}{\cal R}\partial_{i}{\cal X}\,,
\label{f1in}
\end{eqnarray}
where we added and subtracted the quantity $a^{3}q_{1}Q\dot{{\cal R}}^{3}$.
Finally we integrate this term by parts as follows
\begin{eqnarray}
a^{3}q_{1}Q\dot{{\cal R}}^{3} 
& \doteq  & 
-{\cal R}\left[\dot{q}_{1}a^{3}Q\dot{{\cal R}}^{2}
+q_{1}\frac{d}{dt}(a^{3}Q)\dot{{\cal R}}^{2}
+2q_{1}a^{3}Q\dot{{\cal R}}\ddot{{\cal R}}\right]
\nonumber \\
&\doteq & -{\cal R}\dot{{\cal R}}\left[\dot{q}_{1}a^{3}Q\dot{{\cal R}}+q_{1}\frac{d}{dt}(a^{3}Q)\dot{{\cal R}}+2q_{1}a^{3}Q\ddot{{\cal R}}\right]
\doteq -{\cal R}\dot{{\cal R}}\left[2q_{1}\frac{d}{dt}(a^{3}Q\dot{{\cal R}})
-q_{1}\frac{d}{dt}(a^{3}Q)\dot{{\cal R}}+\dot{q}_{1}a^{3}Q\dot{{\cal R}}\right] 
\nonumber \\
& \doteq  & -2q_{1}{\cal R}\dot{{\cal R}}\,\frac{d}{dt}(a^{3}Q\dot{{\cal R}})
+a^{3}[q_{1}(\dot{Q}+3HQ)-Q\dot{q}_{1}]{\cal R}\dot{{\cal R}}^{2}\,.
\end{eqnarray}
Then Eq.~(\ref{f1in}) reads
\begin{eqnarray}
\hspace{-0.9cm} a^{3}f_{1} & \doteq  & a^{3}\left[A_{4}-A_{6}\frac{Q^{2}}{w_{1}^{2}}+Q^{2}\,
\frac{d}{dt}\left(\frac{A_{5}}{w_{1}^{2}}\right)-\frac{3HA_{5}Q^{2}}{w_{1}^{2}}
+q_{1}(\dot{Q}+3HQ)-Q\dot{q}_{1}\right]{\cal R}\dot{{\cal R}}^{2} \nonumber \\
\hspace{-0.9cm}& &+a^{3}\left(A_{1}+A_{3}\frac{Q}{w_{1}}-q_{1}\, Q\right)
\dot{{\cal R}}^{3}-2q_{1}{\cal R}\dot{{\cal R}}\,\frac{d}{dt}(a^{3}Q\dot{{\cal R}})+\frac{a^{3}}{w_{1}}\left(A_{9}\frac{Q}{w_{1}}\right)\dot{{\cal R}}\partial_{i}{\cal R}\partial_{i}{\cal X} \nonumber \\
\hspace{-0.9cm} & &+a^{3}\left[\frac{A_{6}}{w_{1}^{2}}-\frac{d}{dt}\left(\frac{A_{5}}{w_{1}^{2}}\right)+\frac{3HA_{5}}{w_{1}^{2}}\right]{\cal R}(\partial_{ik}{\cal X})(\partial_{ik}{\cal X})-\frac{2A_{5}}{w_{1}^{2}}\frac{d}{dt}(a^{3}Q\dot{{\cal R}})[(\partial_{k}{\cal R})(\partial_{k}{\cal X})-\partial^{-2}\partial_{i}\partial_{j}[(\partial_{i}{\cal R})(\partial_{j}{\cal X})]].
\end{eqnarray}
The value of $q_{1}$ is chosen to match another term
in $af_{2}$, see Eq.~(\ref{q1con}). 
Using the following relation 
\begin{equation}
c(t)\,(\partial_{j}{\cal X})(\partial_{i}{\cal R})(\partial_{ij}{\cal X})
\doteq -c\,(\partial^{2}{\cal R})(\partial{\cal X})^{2}/2\,,
\end{equation}
into the following equality
\begin{eqnarray}
c(t)\,{\cal R}(\partial_{ik}{\cal X})(\partial_{ik}{\cal X}) & \doteq  & -c\,(\partial_{j}{\cal X})[\partial_{i}{\cal R}\partial_{ij}{\cal X}+{\cal R}\partial_{j}\partial^{2}{\cal X}]\doteq 
c\,(\partial^{2}{\cal R})(\partial{\cal X})^{2}/2+c\,(\partial^{2}{\cal X})[{\cal R}(\partial^{2}{\cal X})+(\partial_{i}{\cal R})(\partial_{i}{\cal X})] \nonumber \\
 & \doteq  & c\,(\partial^{2}{\cal R})(\partial{\cal X})^{2}/2+c\, Q^{2}{\cal R}\dot{{\cal R}}^{2}+c\, Q\,\dot{{\cal R}}\,(\partial_{i}{\cal R})(\partial_{i}{\cal X})\,,
\end{eqnarray}
we finally obtain the expression of $a^3 f_1$ given in Eq.~(\ref{eq:f1expr2}).


\end{document}